\pdfoutput=1

\documentclass[11pt,twoside,a4paper,cmspaper,final,collab]{cms-tdr}
\def\svnVersion{444220P}\def\cmsCernNoTag{CERN-EP-2017-168}\def\cmsCernDate{\today}\def\cmsMessage{Published in the Journal of High Energy Physics as \href{http://dx.doi.org/10.1007/JHEP01(2018)054}{\doi{10.1007/JHEP01(2018)054.}}}

\begin{document}\cmsNoteHeader{HIG-17-006}

\hyphenation{had-ron-i-za-tion}
\hyphenation{cal-or-i-me-ter}
\hyphenation{de-vices}
\RCS$Revision: 443592 $
\RCS$HeadURL: svn+ssh://alverson@svn.cern.ch/reps/tdr2/papers/HIG-17-006/trunk/HIG-17-006.tex $
\RCS$Id: HIG-17-006.tex 443592 2018-01-31 16:15:34Z obondu $
\newlength\cmsFigWidth
\ifthenelse{\boolean{cms@external}}{\setlength\cmsFigWidth{0.85\columnwidth}}{\setlength\cmsFigWidth{0.4\textwidth}}
\ifthenelse{\boolean{cms@external}}{\providecommand{\cmsLeft}{top\xspace}}{\providecommand{\cmsLeft}{left\xspace}}
\ifthenelse{\boolean{cms@external}}{\providecommand{\cmsRight}{bottom\xspace}}{\providecommand{\cmsRight}{right\xspace}}
\newlength\cmsTabSkip\setlength{\cmsTabSkip}{2ex}

\providecommand{\PGmmp}{\ensuremath{\mu^\mp}\xspace}
\providecommand{\mll}{\ensuremath{m_{\ell\ell}}\xspace}
\providecommand{\Pj}{\ensuremath{\cmsSymbolFace{j}}\xspace}
\providecommand{\mjj}{\ensuremath{m_{\Pj\Pj}}\xspace}
\providecommand{\PV}{\ensuremath{\mathrm{V}}\xspace}
\providecommand{\PX}{\ensuremath{\text{X}}\xspace}
\providecommand{\mx}{\ensuremath{m_{\text{X}}}\xspace}
\providecommand{\mhh}{\ensuremath{m_{\PH\PH}}\xspace}
\providecommand{\costhetastar}{\ensuremath{\cos \theta^{\mathrm{CS}}_{\PH\PH}}\xspace}

\cmsNoteHeader{HIG-17-006}
\title{Search for resonant and nonresonant Higgs boson pair production in the $\bbbar \ell\nu \ell\nu$ final state in proton-proton collisions at $\sqrt{s} = 13\TeV$}

\date{\today}

\abstract{
Searches for resonant and nonresonant pair-produced Higgs bosons ($\PH\PH$)
decaying respectively into $\ell\nu \ell\nu$, through either $\PW$ or $\Z$ bosons,
and $\bbbar$ are presented.
The analyses are based on a sample of proton-proton collisions at $\sqrt{s} =
13\TeV$, collected by the CMS experiment at the LHC, corresponding to an integrated luminosity
of 35.9\fbinv. Data and predictions from
the standard model are in agreement within uncertainties.
For the standard model
$\PH\PH$ hypothesis, the data exclude at 95\% confidence level a product of the
production cross section and branching fraction larger than 72\unit{fb},
corresponding to 79 times the standard model prediction.
Constraints are placed on different scenarios considering
anomalous couplings, which could affect the rate and kinematics of
$\PH\PH$ production.
Upper limits at 95\% confidence level are set on the production cross section
of narrow-width spin-0 and spin-2 particles decaying to Higgs boson pairs,
the latter produced with minimal gravity-like coupling.
}

\hypersetup{%
pdfauthor={CMS Collaboration},%
pdftitle={Search for resonant and nonresonant Higgs boson pair production in the bblnulnu final state in proton-proton collisions at sqrt(s) = 13 TeV},%
pdfsubject={CMS},%
pdfkeywords={CMS, physics, software, computing}}

\maketitle

\section{Introduction}

The Brout--Englert--Higgs mechanism is a key element of the standard
model (SM) of elementary particles and their interactions, explaining the origin of mass through spontaneous breaking of electroweak symmetry~\cite{Higgs1,Higgs2,Higgs3,Higgs4,Higgs5,Higgs6}.
The discovery of a Higgs boson with a mass $m_\PH$ around
125\GeV by the ATLAS and CMS experiments~\cite{HiggsAtlas, HiggsCms, HiggsCms2}
fixes the value, in the SM, of the self-coupling $\lambda$
in the scalar potential, whose shape is determined by the symmetries of the SM
and the requirement of renormalisability. Direct information on the Higgs
three- and four-point interactions will provide an indication of the scalar
potential structure.

\begin{figure}[!htb]
\centering
\includegraphics[width=0.32\textwidth]{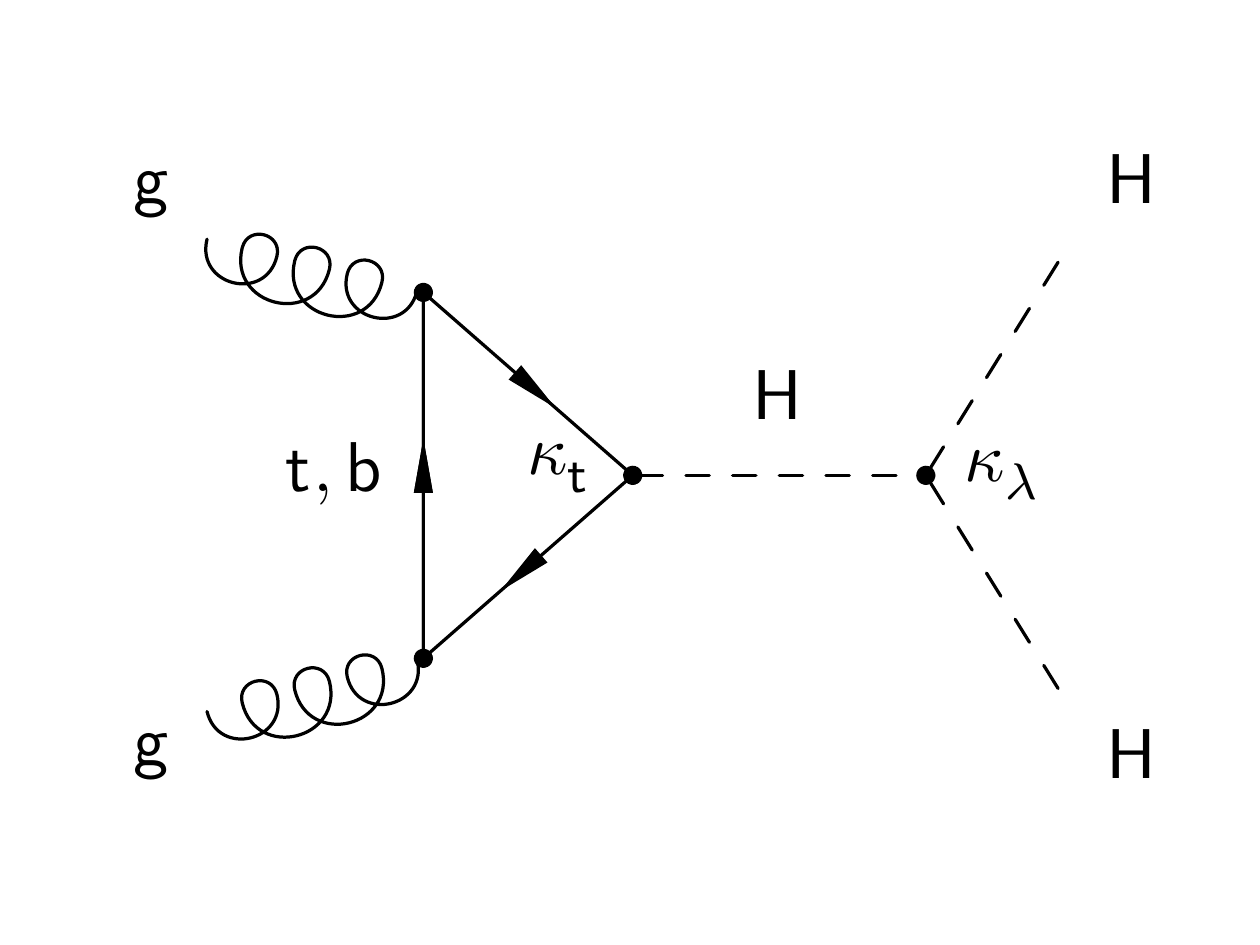}
\includegraphics[width=0.32\textwidth]{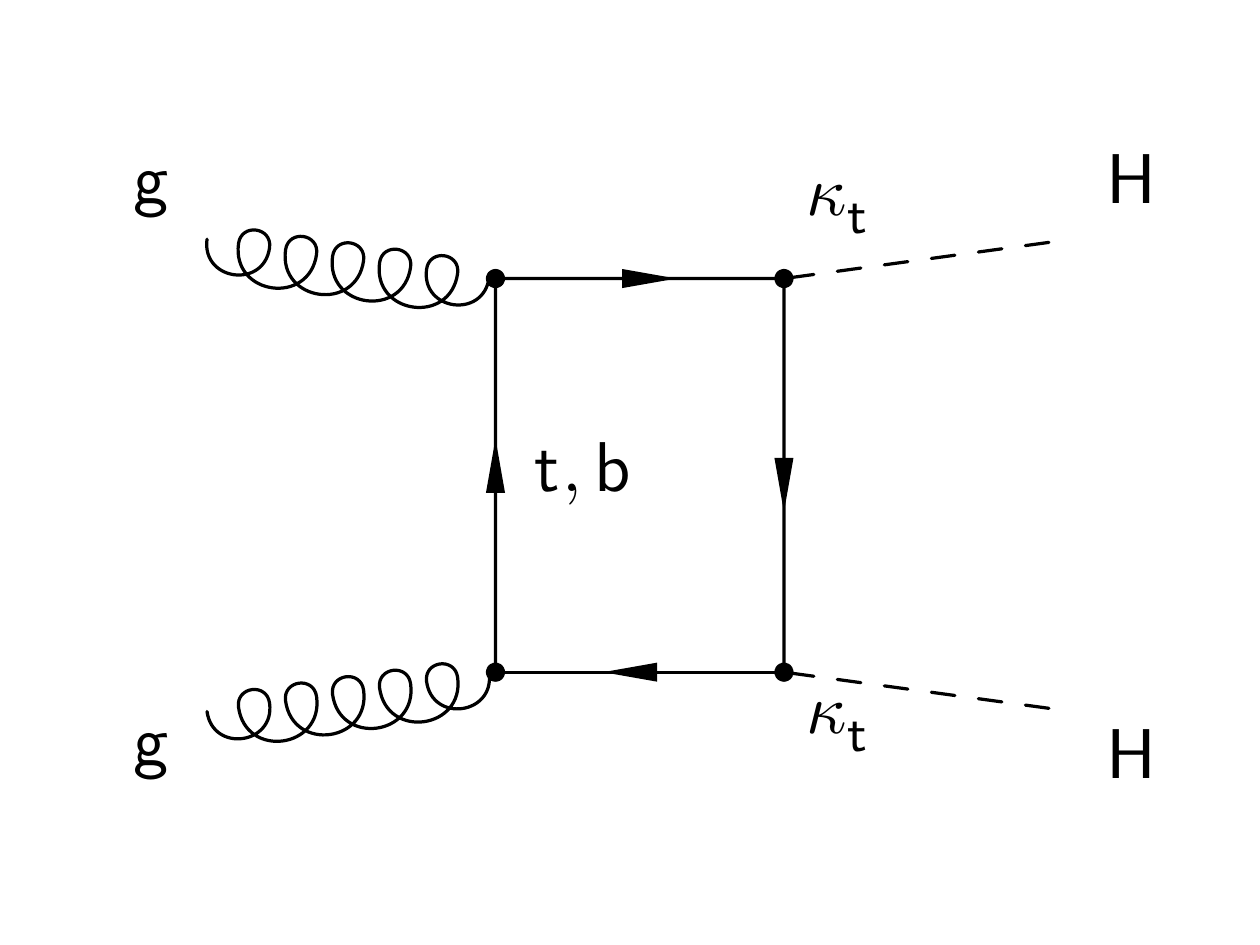}
\caption{
   Feynman diagrams for Higgs boson pair production via gluon fusion in the SM.
   The coupling modifiers for the Higgs boson self-coupling and the top quark Yukawa coupling are denoted by $\kappa_{\lambda}$ and $\kappa_\PQt$, respectively.
}
\label{fig:diagrams}
\end{figure}

Nonresonant Higgs boson pair production ($\PH\PH$) can be used to directly study the Higgs boson
self-coupling. At the CERN LHC, Higgs boson pairs are predominantly produced
through gluon-gluon fusion via two destructively interfering diagrams, shown in Fig.~\ref{fig:diagrams}.
In the SM the destructive interference between these two diagrams
makes the observation of $\PH\PH$ production
extremely challenging, even in the most optimistic scenarios
of energy and integrated luminosity at the future High Luminosity
LHC~\cite{CMS-PAS-FTR-15-002, CMS-PAS-FTR-16-002}.
The SM cross section for $\PH\PH$ production in proton-proton
collisions at $\sqrt{s} = 13$\TeV for a Higgs boson mass of 125\GeV is
$\sigma^{\PH\PH} = 33.5$\unit{fb} at next-to-next-to-leading order (NNLO) in quantum chromodynamics (QCD) for the
gluon-gluon fusion process~\cite{deFlorian:2016spz, deFlorian:2015moa,
  deFlorian:2013jea, Borowka:2016ehy, Dawson:1998py, Grigo:2014jma,
  Grigo:2015dia, Grigo:2013rya, Frederix:2014hta, Maltoni:2014eza}.

Indirect effects at the electroweak scale arising from beyond the standard model (BSM) phenomena at a higher scale
can be parameterised in an effective field theory framework~\cite{Azatov:2015oxa,Goertz:2014qta,Hespel:2014sla} by introducing coupling modifiers for the SM parameters involved in $\PH\PH$ production, namely $\kappa_\lambda = \lambda/\lambda_{\mathrm{SM}}$ for the Higgs boson self-coupling $\lambda$ and $\kappa_\PQt = y_\PQt/y_{\PQt_{\mathrm{SM}}}$ for the top quark Yukawa coupling $y_\PQt$.
Such modifications of the Higgs boson couplings could enhance
Higgs boson pair production to rates observable with the current dataset.
The relevant part of the modified Lagrangian takes the form:
\begin{equation} \label{eq:lag}
{\cal L}_{\PH} =
\frac{1}{2} \partial_{\mu} \PH \, \partial^{\mu} \PH - \frac{1}{2} m_{\PH}^2 \PH^2 -
  {\kappa_{\lambda}} \, \lambda_{\mathrm{SM}} \, v \, \PH^3
- \frac{ m_{\PQt}}{v}(v+   {\kappa_\PQt} \,   \PH ) \,( {\PAQt}_{\text{L}}\PQt_{\mathrm{R}} + \text{h.c.}),
\end{equation}
where $\PH$ is the Higgs boson field, $v$ is the vacuum expectation value of $\PH$, $m_\PQt$ is the top quark mass, $\bar{\PQt}_{\text{L}}$ and $\PQt_{\mathrm{R}}$ are the left- and right-handed top quark fields, and h.c. is the Hermitian conjugate.
The appearance of new contact-like interactions, not considered in this paper, could also result in
an enhancement of $\PH\PH$ production.

Extensions of the scalar sector of the SM postulate the existence of additional
Higgs bosons. An explored scenario is the two-Higgs-doublet model (2HDM)~\cite{Branco:2011iw}, where a
second doublet of complex scalar fields is added to the SM scalar sector Lagrangian.
The generic 2HDM potential has a large number of degrees of freedom,
which can be reduced to six under specific assumptions.
In case the new CP-even state is massive enough (mass larger than $2m_\PH$) it can decay to
a pair of Higgs bosons.
Models inspired by warped extra dimensions~\cite{wed} predict the existence of
new heavy particles that can decay to pairs of Higgs bosons. Examples of such particles are
the radion (spin~0)~\cite{radion1, radion2, radion3, radion4} or the first
Kaluza--Klein (KK) excitation of the graviton (spin~2)~\cite{kk1, kk2}. In
the following, we will use \PX to denote a generic state decaying into
pairs of Higgs bosons.

Searches for Higgs boson pair production have been performed
by the ATLAS and CMS experiments using LHC proton-proton collision data.
These include searches for BSM production as well as more targeted searches for production with SM-like kinematics
in $\sqrt{s}=8$\TeV~\cite{Khachatryan:2140996,Aad:2014yja,Aad:2015uka,Aad:2015xja,Sirunyan:2017tqo}
and 13\TeV data~\cite{Aaboud:2016xco,Sirunyan:2017djm}.

This paper reports on a search for Higgs boson pair production,
$\PH\PH$, and resonant Higgs boson pair production, $\PX \to \PH\PH$, where one of the $\PH$ decays into $\bbbar$, and the
other into $\PZ(\ell\ell)\PZ(\PGn\PGn)$ or
$\PW(\ell\PGn)\PW(\ell\PGn)$, where $\ell$ is either an electron, a
muon, or a tau lepton that decays leptonically.
The search is based on LHC proton-proton collision data at $\sqrt{s}=13$\TeV collected by the CMS experiment,
corresponding to an integrated luminosity of 35.9\fbinv. The analysis
focuses on the invariant mass distribution of the b jet pairs,
searching for a resonant-like excess compatible with the Higgs boson mass, in
combination with an artificial neural network discriminator based on kinematic
information. The dominant background is \ttbar production, with
smaller contributions from Drell--Yan (DY) and single top quark
production.

\section{The CMS detector}

The central feature of the CMS apparatus is a superconducting solenoid
of 6\unit{m} internal diameter, providing a magnetic field of 3.8\unit{T}.
Within the solenoid volume are a silicon pixel and strip tracker,
a lead tungstate crystal electromagnetic calorimeter (ECAL),
and a brass and scintillator hadron calorimeter,
each composed of a barrel and two endcap sections.
Forward calorimeters extend the pseudorapidity coverage
provided by the barrel and endcap detectors.
Muons are detected in gas-ionisation chambers embedded in
the steel flux-return yoke outside the solenoid.
A more detailed description of the CMS detector,
together with a definition of the coordinate system used
and the relevant kinematic variables, can be found in Ref.~\cite{cms}.

\section{Event simulation}

The main background processes, in decreasing order of importance, are \ttbar, DY, and
single top quark production. Diboson, triboson, \ttbar{}V
and SM single Higgs boson production
are also considered.
Other contributions, such as \PW{}+jets or QCD multijet
events with jets misidentified as leptons, are negligible due to the dilepton selection described in the next section.
The dominant contribution, especially in the $\Pepm\PGmmp$ channel, arises
from $\ttbar$ production yielding the same final state as the signal
process (two b quark jets, two leptons, and two neutrinos)
when both \PW{} bosons from top quark decays further decay leptonically.

Background simulation samples have been generated at next-to-leading order (NLO) in QCD using
\POWHEG~2~\cite{Nason:2004rx,Frixione:2007vw,Alioli:2010xd,Powheg_tt,Powheg_st}, and
\MGvATNLO versions~2.2.2.0 and~2.3.2.2~\cite{Alwall:2014hca} with FxFx merging~\cite{Frederix:2012ps} and \textsc{MadSpin}~\cite{Artoisenet:2012st}.
The signal samples of gluon fusion
production of two Higgs bosons, and of spin-0 or spin-2 narrow resonances decaying
into two Higgs bosons, have been generated at leading order (LO) in QCD using \MGvATNLO
version~2.2.2.0.
The spin-2 narrow resonance is produced as a KK-graviton with minimal coupling~\cite{Oliveira:2014kla}, leading to spin projection $\pm$2 on the beam axis.
The mass of the Higgs boson has been fixed to
125\GeV~\cite{Aad:2015zhl}, and its branching fractions to those in the SM. One of the Higgs bosons is required to decay into a pair of b quarks, while the second one is required to decay to final states containing two leptons and two neutrinos of any flavour. This
implies that the signal samples contain both $\PH \to \PZ(\ell\ell)\PZ(\PGn\PGn)$ and
$\PH \to \PW(\ell\PGn)\PW(\ell\PGn)$ decay chains, leading to a total branching fraction
$\mathcal{B}(\PH\PH \to \bbbar\PV\PV \to \bbbar\ell\PGn\ell\PGn)$ of~2.7\%~\cite{deFlorian:2016spz}.
The event generators used for both signal and background samples are interfaced with \PYTHIA~8.212~\cite{Sjostrand:2006za,Sjostrand:2007gs} for simulation of the parton showering, hadronisation, and underlying event using the CUETP8M1 tune~\cite{Khachatryan:2110213}. The NNPDF~3.0~\cite{Ball:2014uwa} LO and NLO Parton Distribution Functions (PDF) are used.

For all processes, the detector response is simulated using a detailed description of the CMS
apparatus, based on the \GEANTfour package~\cite{Agostinelli2003250}. Additional $\Pp\Pp$ interactions in the same and in the neighbouring bunch crossings (pileup) are generated with \PYTHIA and overlapped with the simulated events of interest to reproduce the pileup measured in data.

All background processes are normalised to their most accurate theoretical cross sections. The \ttbar, DY, single top quark and $\PW^+ \PW^-$ samples are normalised to NNLO precision in QCD~\cite{Czakon:2011xx,Li:2012wna,Kidonakis:2013zqa,Gehrmann:2014fva}, while remaining diboson, triboson and $\ttbar \PV$ processes are normalised to NLO precision in QCD~\cite{Campbell:2011bn,Alwall:2014hca}. The single Higgs boson production cross section is computed at the NNLO precision of QCD corrections and NLO precision of electroweak corrections~\cite{deFlorian:2016spz}.

\section{Event selection and background predictions}

Data are collected with a set of dilepton triggers. The $\pt$ thresholds applied by the triggers are asymmetric and channel-dependent, and vary from 17 to 23\GeV for the leading leptons and from 8 to 12\GeV for the subleading leptons. Trigger efficiencies are measured with a ``tag-and-probe'' technique~\cite{Chatrchyan2011} as a function of lepton $\pt$ and $\eta$ in a data control region consisting of $\PZ \to \ell\ell$ events.

Events with two oppositely charged leptons
(\Pep\Pem, \PGmp{}\PGmm, \Pepm{}\PGmmp) are selected
using asymmetric $\pt$ requirements, chosen to be above the corresponding trigger thresholds, for leading and subleading leptons of 25\GeV
and 15\GeV for $\Pe \Pe$ and $\PGm \Pe$ events, 20\GeV and 10\GeV for $\PGm\PGm$ events,
and 25\GeV and 10\GeV for $\Pe\PGm$ events.
Electrons in the pseudorapidity range
$\abs{\eta} < 2.5$ and muons in the range $\abs{\eta} < 2.4$ are considered.

Electrons, reconstructed by associating tracks with ECAL clusters,
are identified by a sequential selection using information
on the cluster shape in the ECAL, track quality, and the matching
between the track and the ECAL cluster. Additionally, electrons from
photon conversions are rejected~\cite{Khachatryan:2015hwa}. Muons are reconstructed from tracks
found in the muon system, associated with tracks in the silicon tracking
detectors. They are identified based on the quality of the track fit
and the number of associated hits in the different tracking detectors~\cite{Chatrchyan:2012xi}.
The lepton isolation, defined as the scalar $\pt$ sum of all particle candidates in a cone around the
lepton, excluding the lepton itself, divided by the lepton $\pt$, is
required to be $<$0.04 for electrons (with a cone of radius $\Delta R = \sqrt{\smash[b]{(\Delta \phi)^2 + (\Delta \eta)^2}} = 0.3$) and $<$0.15 for muons (with a cone of radius $\Delta R = 0.4$). Lepton
identification and isolation efficiencies in the simulation are corrected for
residual differences with respect to data. These corrections are measured in a data sample, enriched in $\PZ \to \ell\ell$ events, using a ``tag-and-probe'' method and are parameterised as a function of lepton $\pt$ and $\eta$.

Jets are reconstructed using a particle flow (PF) technique~\cite{Sirunyan:2017ulk}.
PF candidates are clustered to form jets using the anti-\kt
clustering algorithm~\cite{Cacciari:2008gp} with a distance parameter of 0.4, implemented in the \FASTJET
package~\cite{Cacciari:2011ma}. Jet energies are corrected for residual
nonuniformity and nonlinearity of the detector response~\cite{Khachatryan:2016kdb}. Jets are required to have
$\pt > 20$\GeV, $\abs{\eta} < 2.4$, and be separated from identified leptons
by a distance of $\Delta R > 0.3$.
The missing transverse momentum vector, defined as the projection onto the transverse plane relative to the beam axis, of the negative vector sum of the momenta of all PF candidates, is referred to as $\ptvecmiss$~\cite{metPF,CMS-PAS-JME-16-004}. Its magnitude is denoted
by $\ptmiss$. Corrections to the jet energies are propagated to
$\ptvecmiss$.

The reconstructed vertex with the largest value of summed object $\pt^2$ is taken to be the primary $\Pp\Pp$ interaction vertex, considering the objects returned by a clustering algorithm applied to all charged tracks associated with the vertex, plus the corresponding associated $\ptvecmiss$.

The combined multivariate algorithm~\cite{CMSbtag,CMSbtag_Run2} is used to identify jets originating from b quarks.
Jets are considered as b tagged if they pass the medium
working point of the algorithm, which provides around 70\% efficiency
with a mistag rate less than 1\%. Correction factors are applied in the simulation
to the selected jets to account for the different response
of the combined multivariate algorithm between data and simulation~\cite{CMSbtag_Run2}.
Among all possible dijet combinations fulfilling the previous criteria, we select
the two jets with the highest combined multivariate algorithm outputs.

\begin{figure}[!htb]
  \centering
    \includegraphics[width=0.45\textwidth]{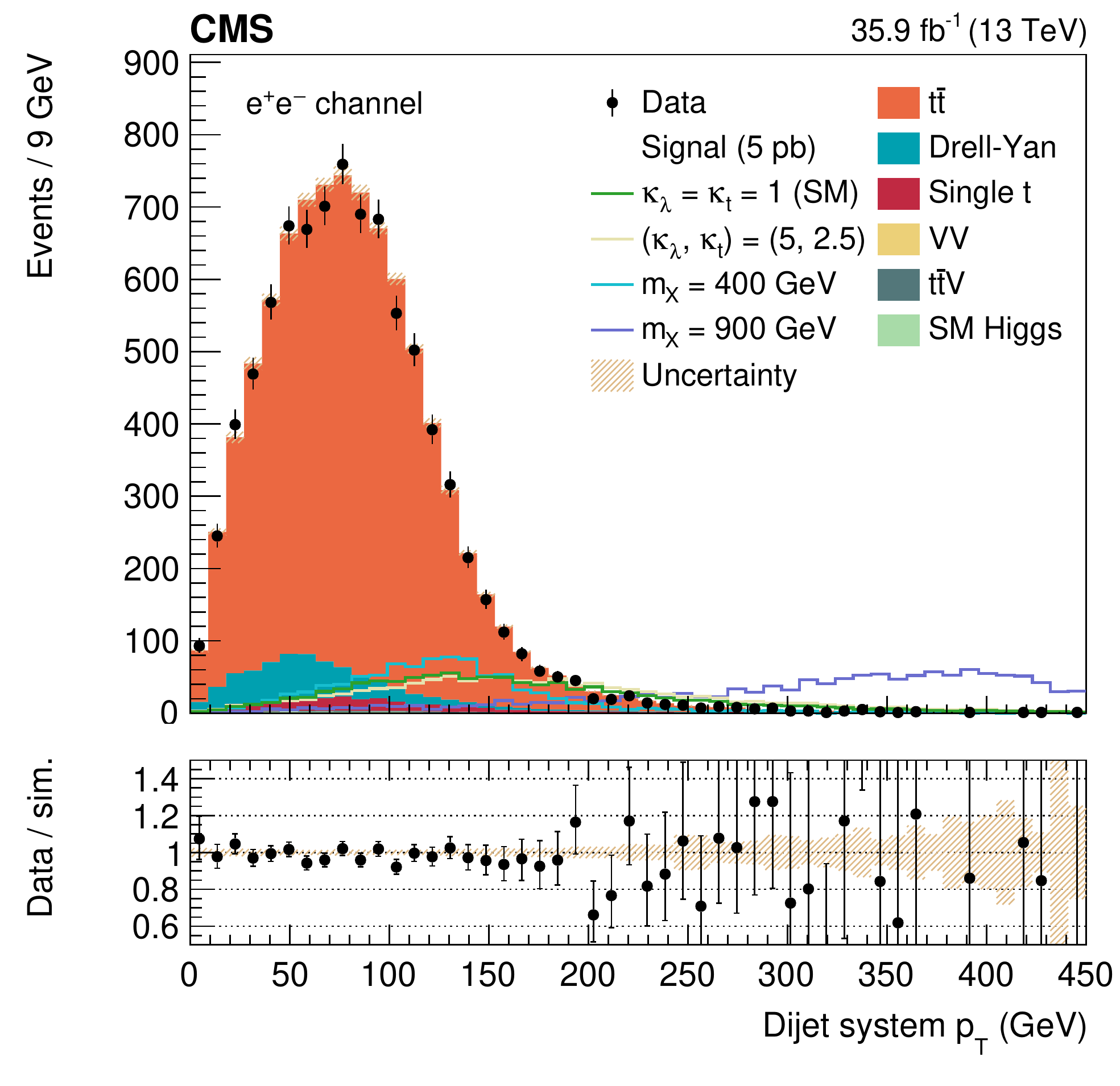}
    \includegraphics[width=0.45\textwidth]{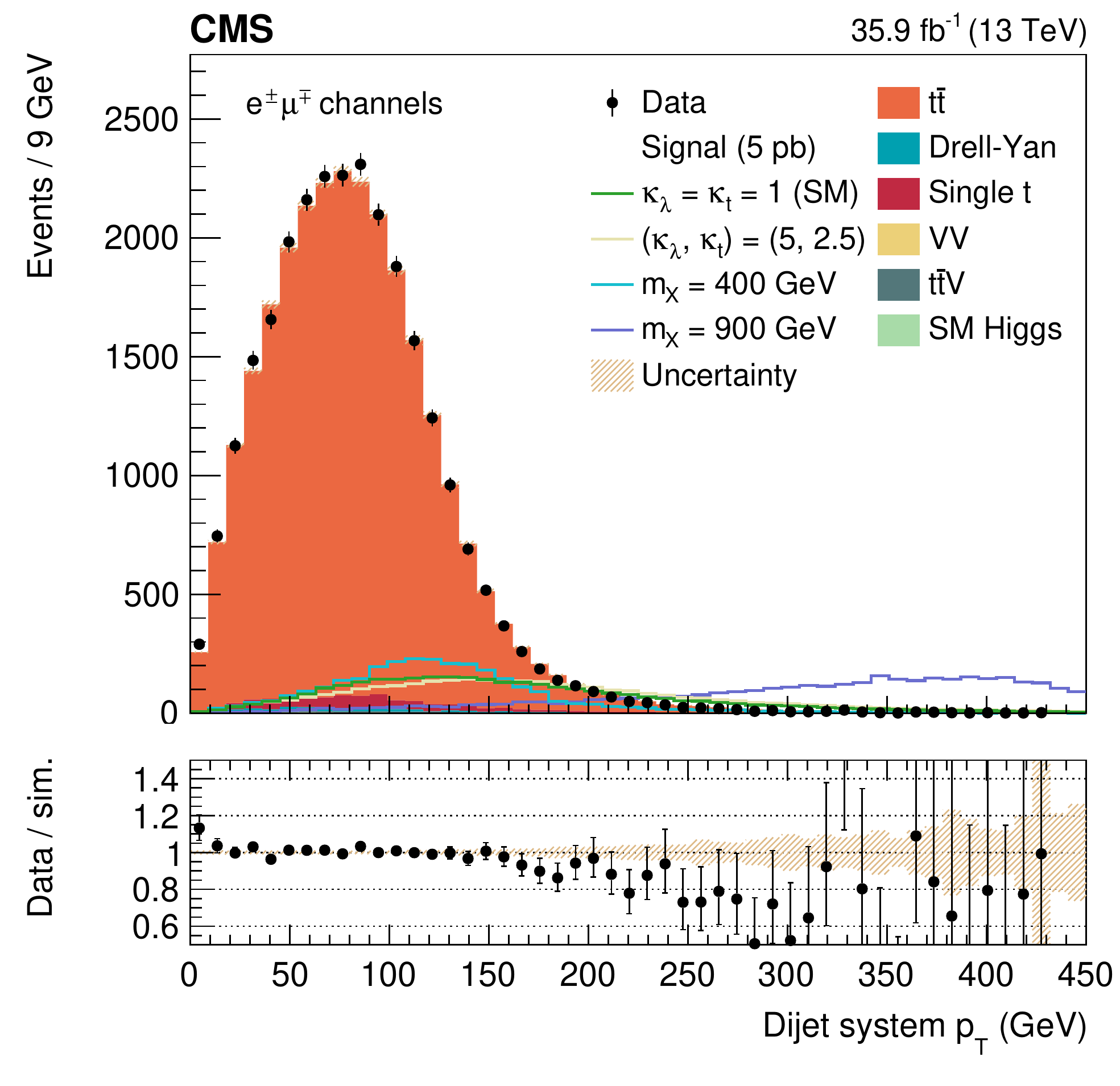}
    \includegraphics[width=0.45\textwidth]{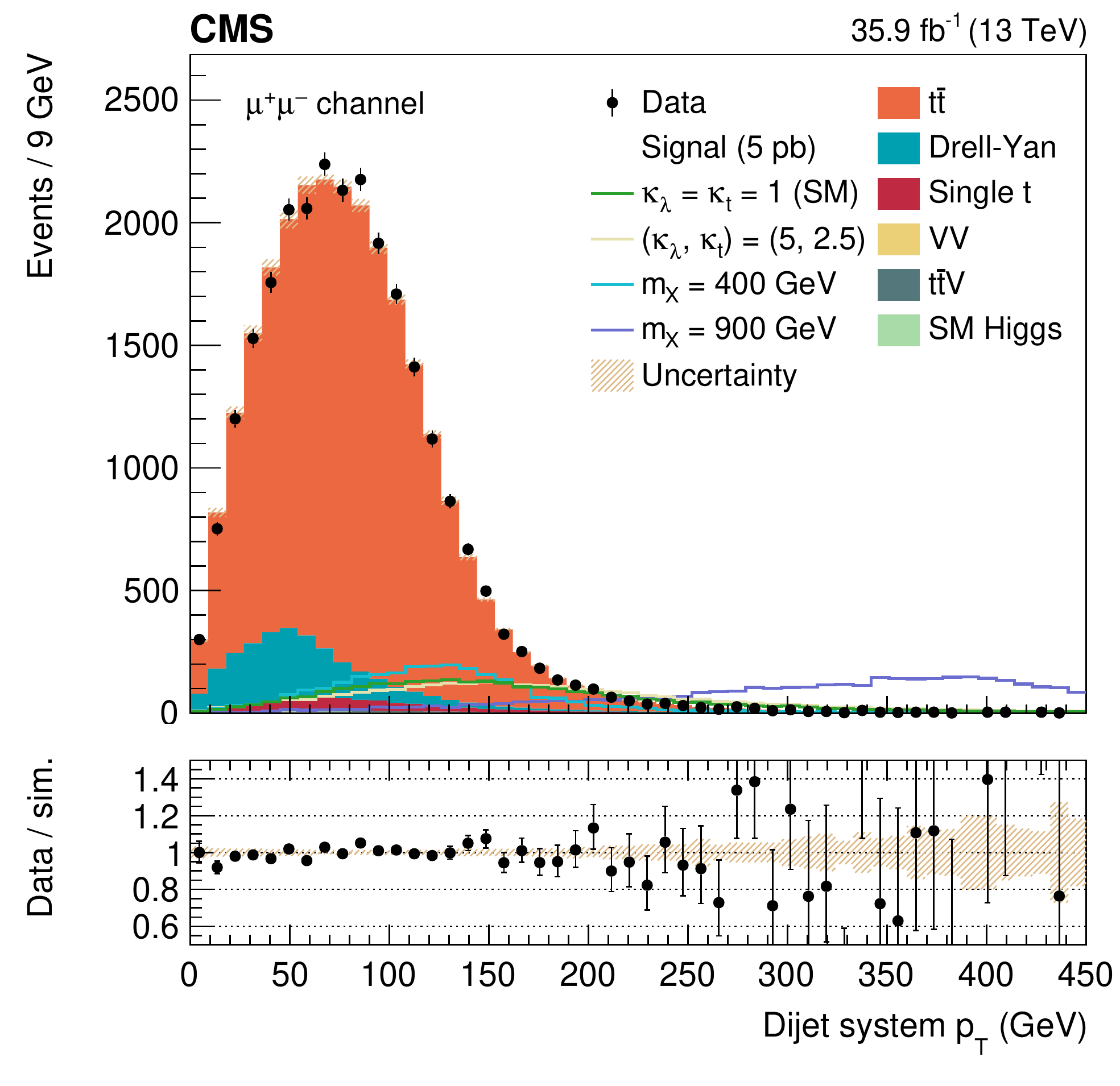}
    \caption{
    The dijet \pt distributions in data and simulated events after requiring two leptons,
    two b-tagged jets, and $12 < \mll < m_{\PZ} - 15$\GeV, for $\Pep\Pem$ (top left), $\Pepm{}\PGmmp$ (top right),
    and $\PGmp{}\PGmm$ (bottom) events. The various signal hypotheses displayed have been scaled to a
    cross section of 5\unit{pb} for display purposes. Error bars indicate statistical uncertainties, while shaded bands show post-fit systematic uncertainties.
    }
    \label{fig:jet_pt}
\end{figure}

After the final object selection consisting of two opposite sign leptons and two
b-tagged jets, a requirement of $12 < \mll < m_{\PZ} - 15$\GeV is applied to suppress quarkonia resonances and jets misidentified as leptons, and to remove
the large background at the \PZ boson peak as well as the high-$\mll$ tail of the DY and $\ttbar$ processes. This requirement has a negligible impact on signal events where one $\PH$ decays as $\PH \to \PW(\ell\PGn)\PW(\ell\PGn)$, and removes only the portion of $\PH \to \PZ(\ell\ell)\PZ(\PGn\PGn)$ decays with on-shell $\PZ(\ell\ell)$ legs.
Figure~\ref{fig:jet_pt} shows the dijet \pt for data
and simulated events after requiring the selection criteria described in this section.

All the background processes are estimated from simulation,
with the exception of DY production in the
$\Pep\Pem$ and $\PGmp\PGmm$ channels.
The DY contribution in the $\Pepm\PGmmp$ channels is
almost negligible, and is taken from simulation.

The contribution of the DY process in the analysis selection is estimated from a data sample enriched in DY plus jets events.
The estimate is performed by requiring all the selection criteria described above, except for the b tagging requirements. The resulting dataset is corrected with weights to represent the DY contribution in the full selection.
The weights are a function of kinematic variables and are
tuned to reproduce the effect of applying the b tagging requirements on the DY process. They account for the following features:
\begin{itemize}
    \item The b tagging efficiencies are not constant and depend on jet kinematics. Moreover, this dependency is different for $\PQb$-, $\PQc$- or light-flavour jets.
    \item The relative contributions of DY plus two jets of flavours $k$ and $l$, where $k, l = \PQb, \PQc, \text{or light-flavour}$, to the DY plus two jets process are not constant throughout the phase-space. Modelling the effect of b tagging requires to parameterise the fractions $F_{kl}$ of jets with flavours $k$ and $l$ as a function of event kinematics.
\end{itemize}

We compute the weights as:
\begin{equation}
    \label{eq:1}
    W_{\text{sim}} = \sum_{k,l=\PQb, \PQc, \text{light-flavour}} \, F_{kl}(\mathrm{BDT}) \, \epsilon_{k}(\pt^{\Pj_1}, \eta^{\Pj_1}) \, \epsilon_{l}(\pt^{\Pj_2}, \eta^{\Pj_2})\text{,}
\end{equation}
where $\epsilon_{k}$ and $\epsilon_{l}$ are the b tagging
efficiencies for $k$- and $l$-flavour jets calculated
from simulation as a function of \pt and $\eta$ of the jets and corrected for differences between data and simulation, and $\Pj_1$ and $\Pj_2$ denote the two $\pt$-ordered jets selected according to the above requirements.
The expected fractions of jets with flavours $k$ and $l$ in the dataset are denoted by $F_{kl}$ and are parameterised as a function of the output
value of a Boosted Decision Tree (BDT)~\cite{Hocker:2007ht}. The indices $k$ and $l$ refer to the assumed flavour of $\Pj_1$ and $\Pj_2$, respectively.
The flavour fractions $F_{kl}$ are estimated from a simulated DY sample.
Their dependency on the BDT output value accounts for the different kinematical behaviours of heavy- or light-flavour associated DY processes, effectively reducing the dimensionality of the phase-space to a single variable. The BDT is trained to discriminate
DY+$\PQb \PAQb, \PQc \PAQc$ from other DY associated production processes
using the following input variables: $\pt^{\Pj_1}$,
$\pt^{\Pj_2}$, $\eta^{\Pj_1}$, $\eta^{\Pj_2}$,
$\pt^{\Pj\Pj}$, $\pt^{\ell\ell}$,
$\eta^{\ell\ell}$, $\Delta \phi(\ell\ell, \ptvecmiss)$ (defined as
the $\Delta \phi$ between the dilepton system and $\ptvecmiss$), number of jets,
and $H_{\mathrm{T}}$ defined as the scalar sum of the transverse
momentum of all selected leptons and jets.

Beside DY, the data sample without b tagging requirements contains small
contributions from other backgrounds such as $\ttbar$, single top
quark and diboson production. Hence, the same reweighting procedure is
applied to simulated samples for these processes, and the result is
subtracted from the weighted data to define the estimate of the DY background in the analysis region.

The method is validated both in simulation and in two data control regions requiring either
$\mll > m_{\PZ} - 15$\GeV or $\abs{\mll - m_{\PZ}} < 15$\GeV. The
predicted DY distributions are in agreement with data and simulation
within the uncertainties of the method, described in
section~\ref{sec:systematics}.

\section{Parameterised multivariate discriminators for signal extraction}
\label{sec:optim}
Deep neural network (DNN) discriminators, based on the Keras library~\cite{chollet2015}, are used to improve the signal-to-background
separation. As the dominant background process ($\ttbar$ production) is irreducible, the DNNs rely on information related
to event kinematics. The variables provided as input to the DNNs exploit the presence in the signal of two Higgs
bosons decaying into two b jets on the one hand, and two leptons and two neutrinos on the other hand,
which results in different kinematics for the dilepton and dijet systems between signal and
background processes. The variables used as input are: \mll,
$\Delta R_{\ell\ell}$, $\Delta R_{\Pj\Pj}$, $\Delta \phi (\ell\ell, \Pj\Pj)$ (defined as
the $\Delta \phi$ between the dijet and the dilepton systems), $\pt^{\ell\ell}$, $\pt^{\Pj\Pj}$,
min$\left(\Delta R_{\Pj \ell}\right)$, and $m_\text{T} = \sqrt{\smash[b]{2 \pt^{\ell\ell} \ptmiss [1 - \cos \Delta
\phi(\ell\ell, \ptvecmiss)]}}$.

\begin{figure}[!htb]
  \centering
    \includegraphics[width=0.55\textwidth]{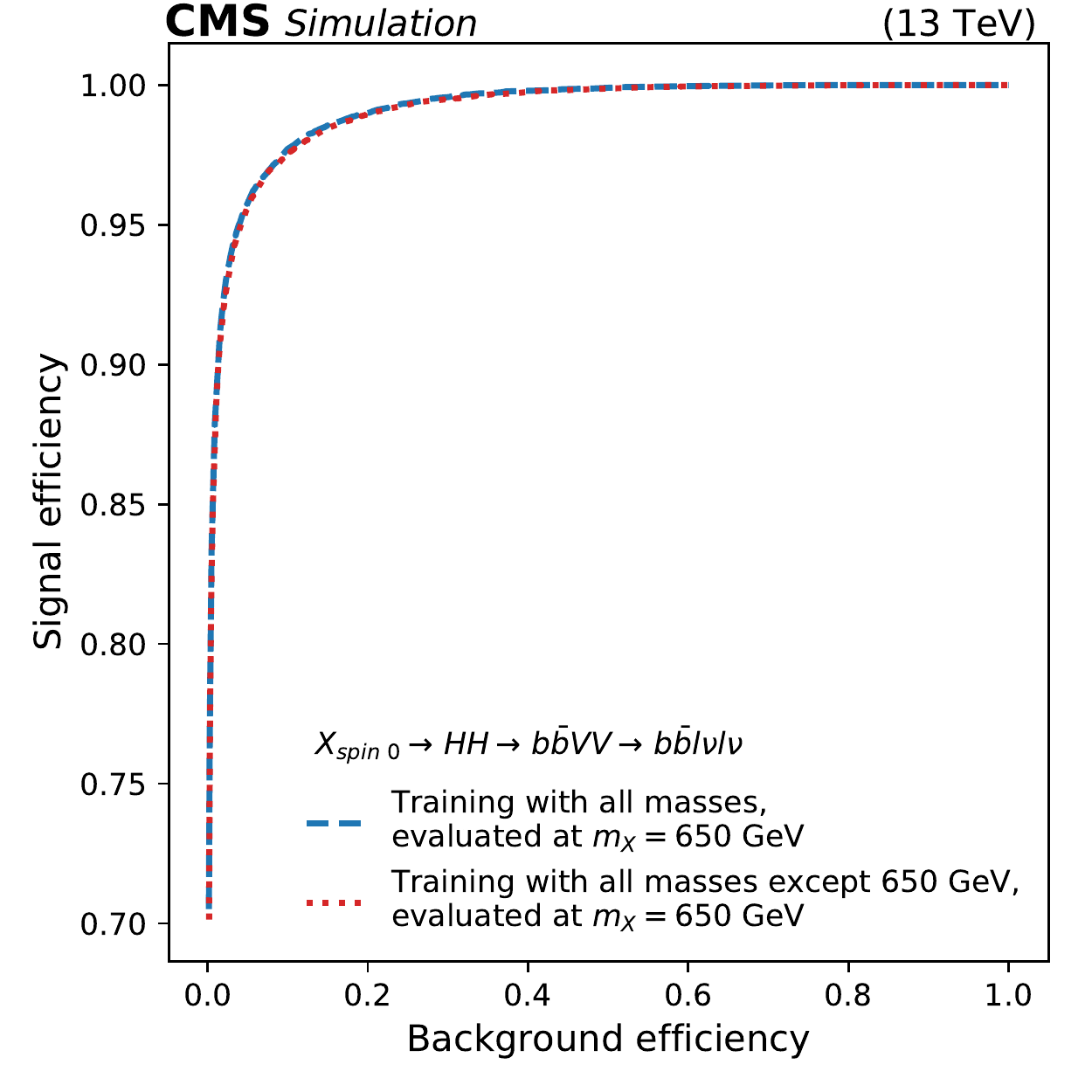}
    \caption{
    Performance of the parameterised DNN for the resonant search, shown as the selection efficiency for the $\mx=$~650~GeV signal as a function of the selection efficiency for the background (ROC curve), for the combined $\Pep\Pem$, $\PGmp\PGmm$ and $\Pepm\PGmmp$ channels. The dashed line corresponds to the DNN used in the analysis, trained on all available signal samples, and evaluated at $\mx=$~650~GeV. The dotted line shows an alternative DNN trained using all signal samples except for $\mx=$~650~GeV, and evaluated at $\mx=$~650~GeV. Both curves overlap, indicating that the parameterised DNN is able to generalise to cases not seen during the training phase by interpolating the signal behaviour from nearby $\mx$ points.
    }
    \label{fig:dnn_roc}
\end{figure}

The DNNs utilise a parameterised machine learning
technique~\cite{Baldi:2016} in order to ensure optimal sensitivity on
the full range of signal hypotheses considered in these searches. In
this approach, one or more physics parameters describing the wider
scope of the problem, as for example the mass of the resonance in the
resonant search case, are provided as input to the DNNs, in addition
to reconstructed quantities. The parameterised network is able to
perform as well as individual networks trained on specific hypotheses
(parameter values) while requiring only a single training, and
provides a smooth interpolation to cases not seen during the training
phase, as shown by Fig. \ref{fig:dnn_roc}. Two parameterised DNNs are trained: one for the resonant and
one for the nonresonant search. In the first case, the set of
parameters are the masses of the resonance, providing 13 values for
the network training ($\mx = 260$, 270, 300, 350, 400, 450, 500, 550,
600, 650, 750, 800, 900\GeV), and a discrete variable indicating the
dilepton flavour channel: same flavour ($\Pep\Pem$ and $\PGmp\PGmm$)
or different flavour ($\Pepm\PGmmp$). In the second case, the
parameters are $\kappa_{\lambda}$ and $\kappa_{\PQt}$, providing 32
combinations of those for the network training ($\kappa_{\lambda} =
-20$, -5, 0, 1, 2.4, 3.8, 5, 20 and $\kappa_{\PQt} = 0.5$, 1, 1.75,
2.5), and the dilepton flavour channel variable as in the resonant
case.

\begin{figure}[!htb]
  \centering
    \includegraphics[width=0.45\textwidth]{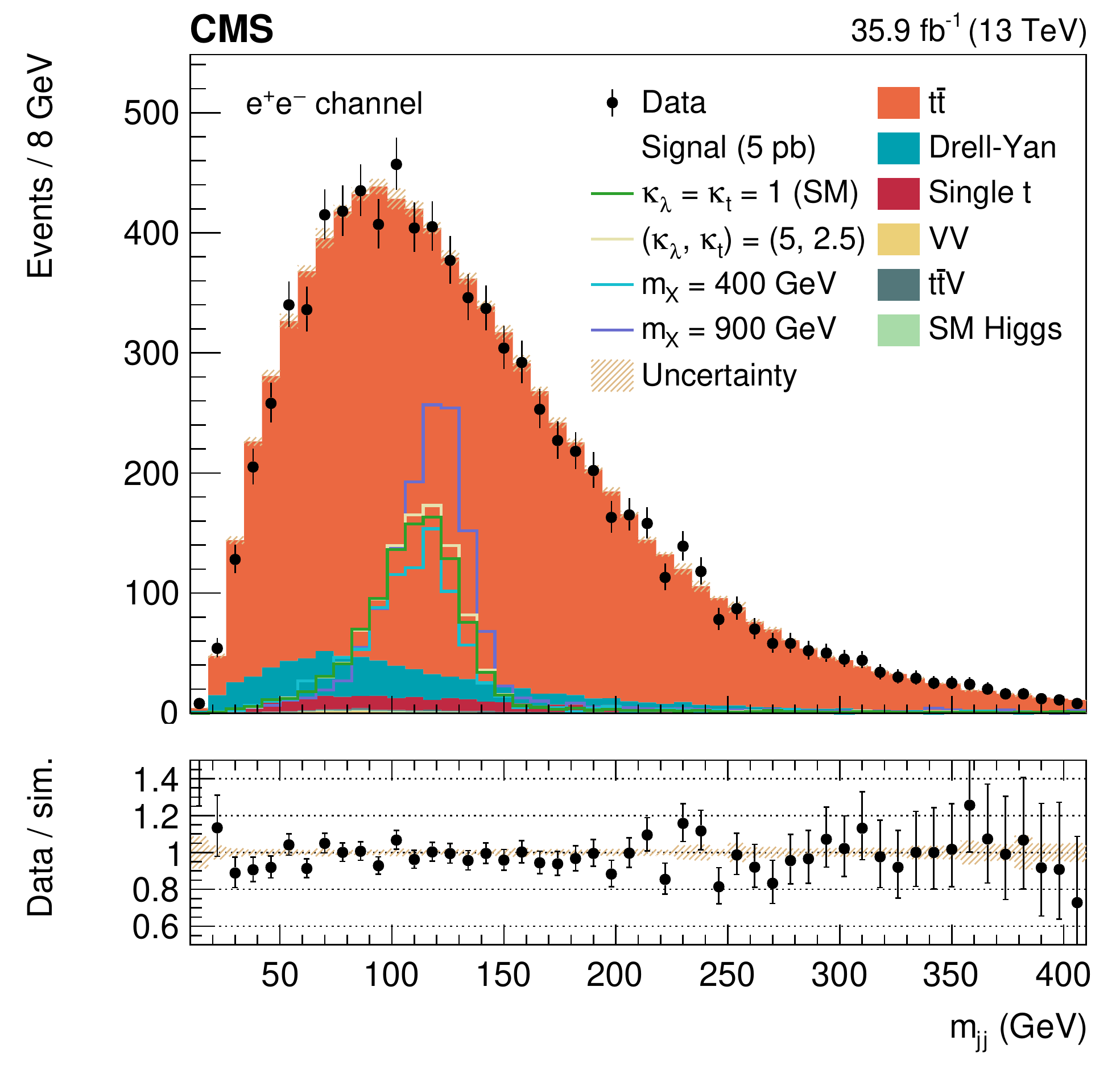}
    \includegraphics[width=0.45\textwidth]{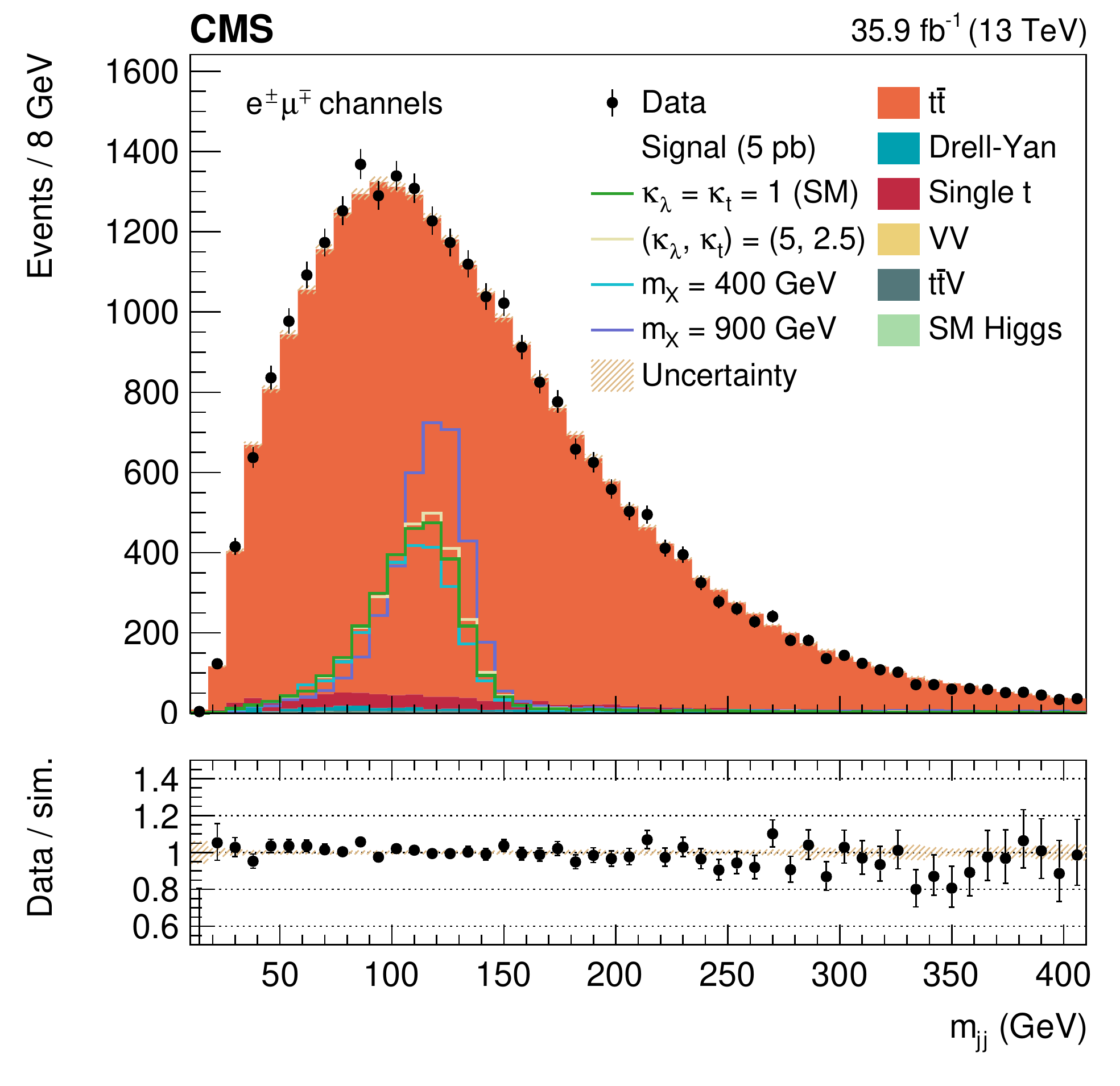}
    \includegraphics[width=0.45\textwidth]{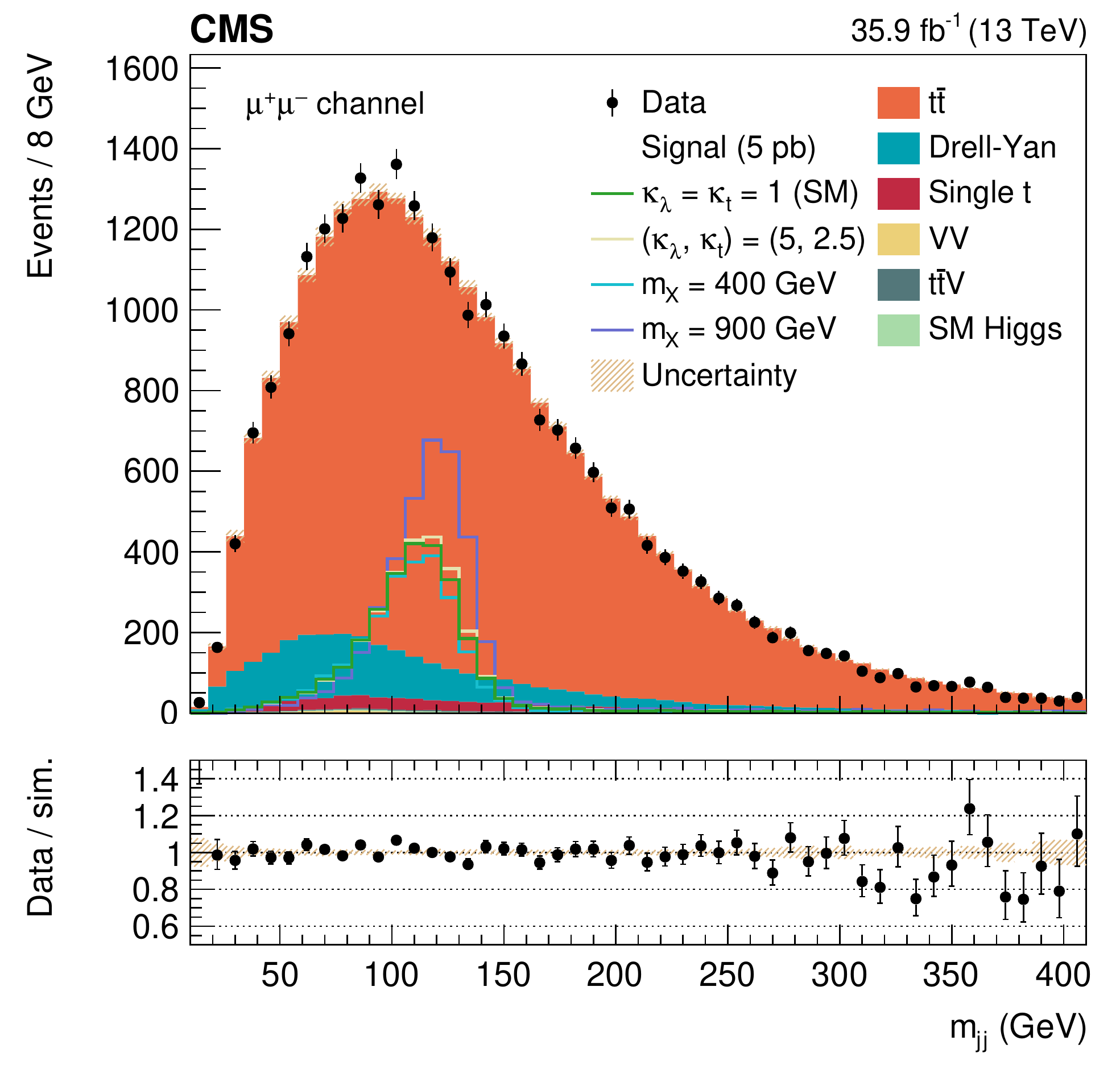}
    \caption{
    The \mjj distribution in data and simulated events
    after requiring all selection criteria in the $\Pep\Pem$ (top left), $\Pepm{}\PGmmp$ (top right),
    and $\PGmp{}\PGmm$ (bottom) channels. The various signal hypotheses displayed have been scaled to a
    cross section of 5\unit{pb} for display purposes. Error bars indicate statistical uncertainties, while shaded bands show post-fit systematic uncertainties.
     }
    \label{fig:mjj}
\end{figure}

\begin{figure}[!htb]
  \centering
    \includegraphics[width=0.45\textwidth]{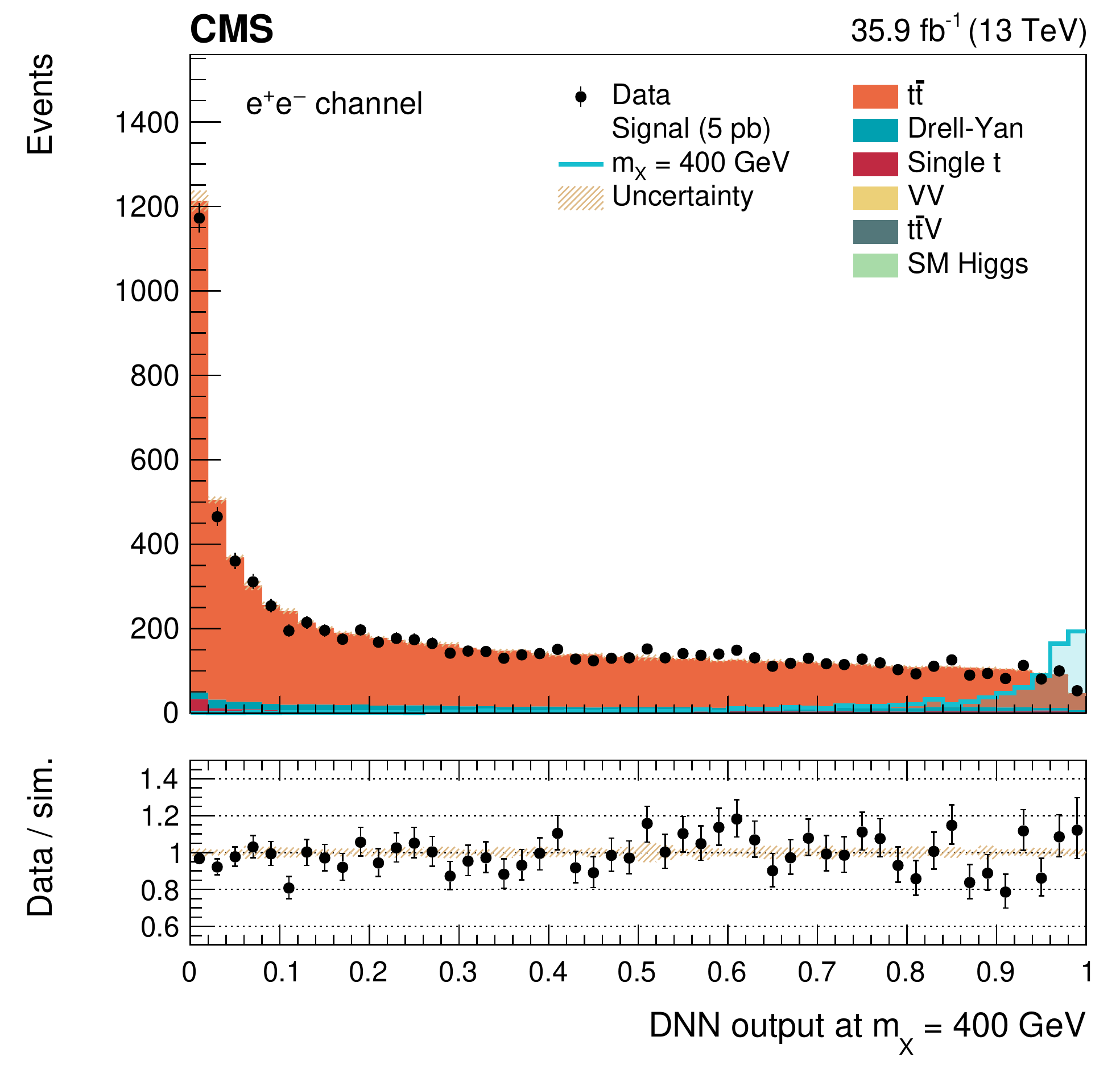}
    \includegraphics[width=0.45\textwidth]{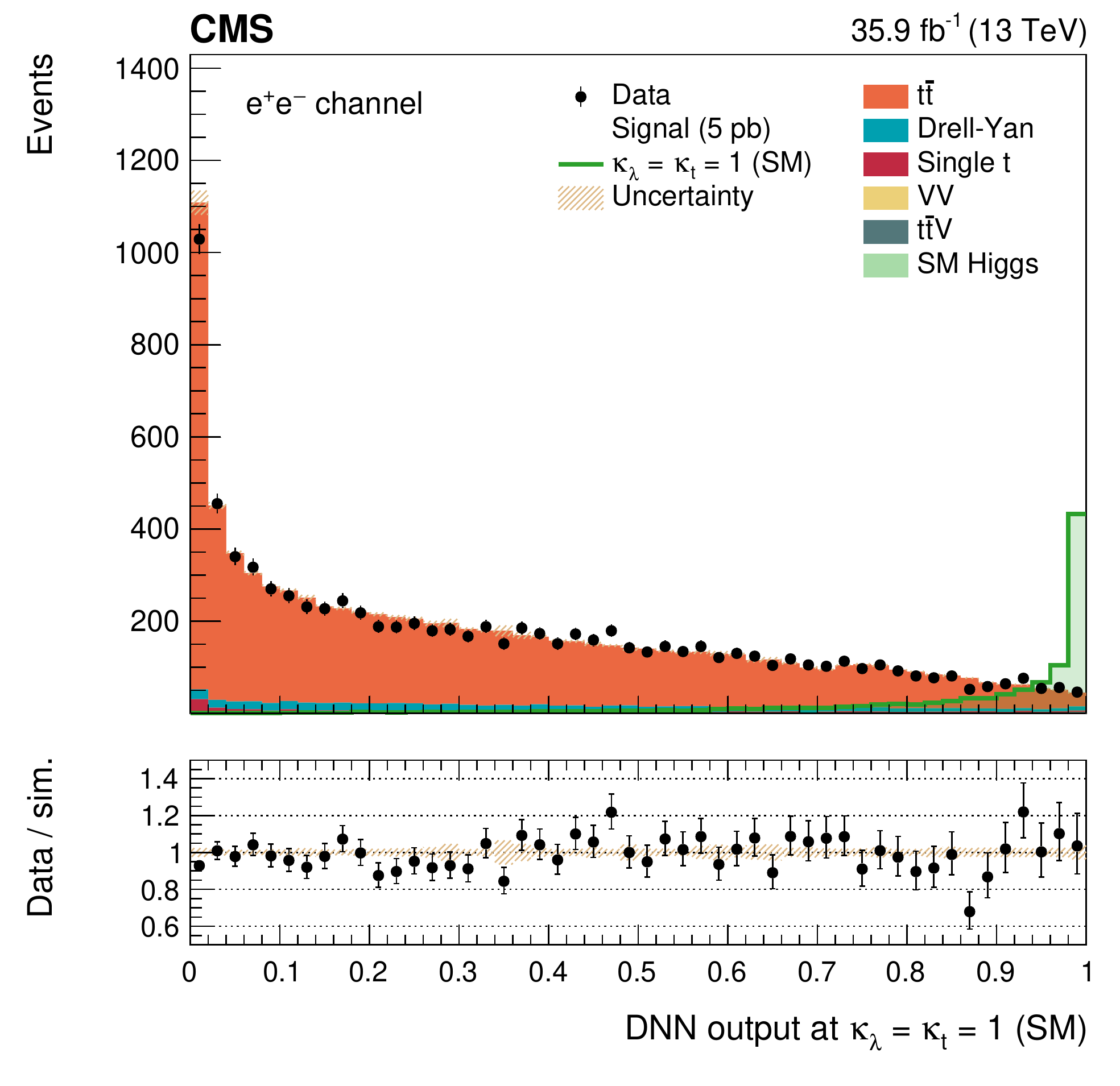}
    \includegraphics[width=0.45\textwidth]{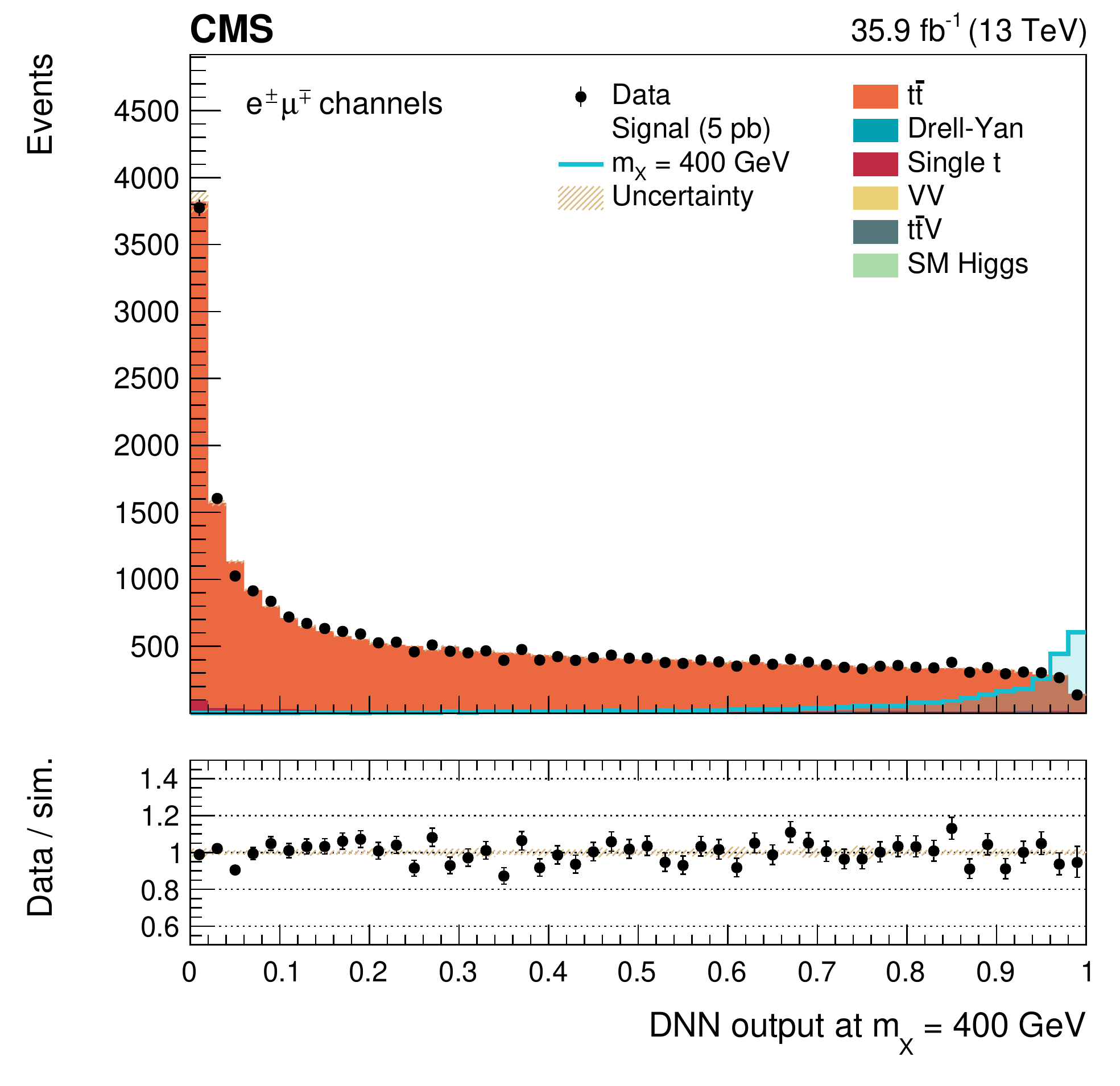}
    \includegraphics[width=0.45\textwidth]{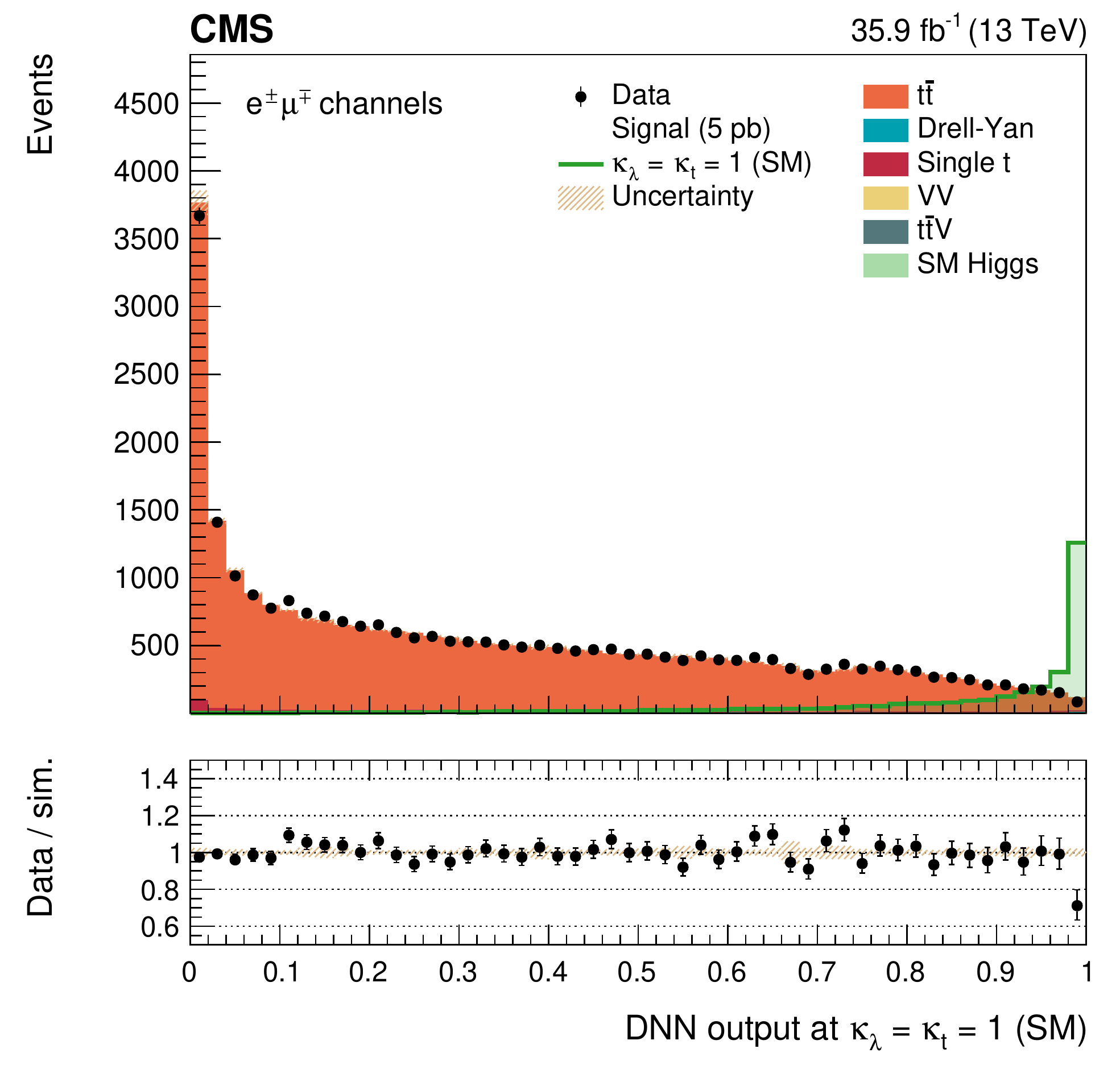}
    \includegraphics[width=0.45\textwidth]{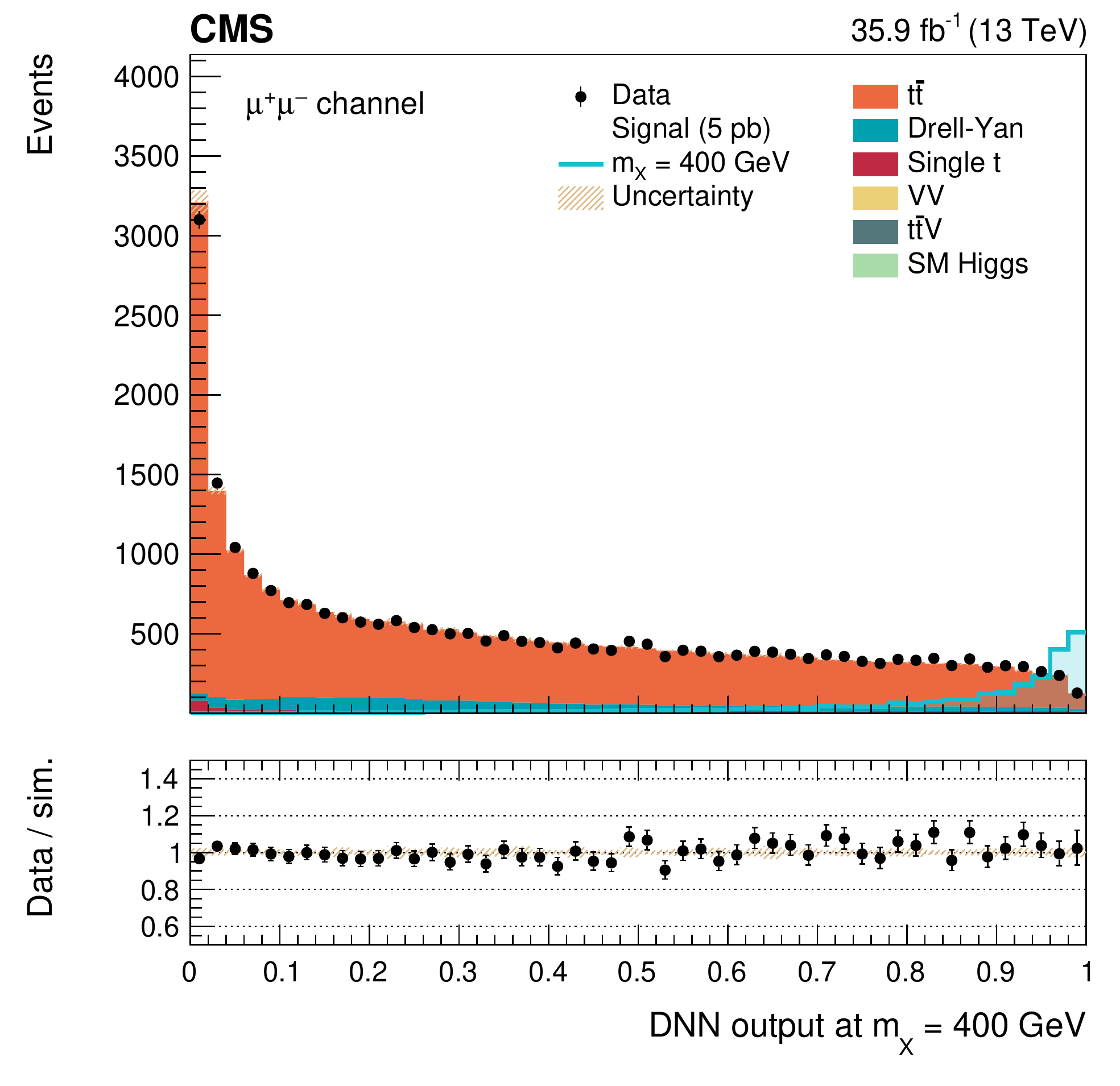}
    \includegraphics[width=0.45\textwidth]{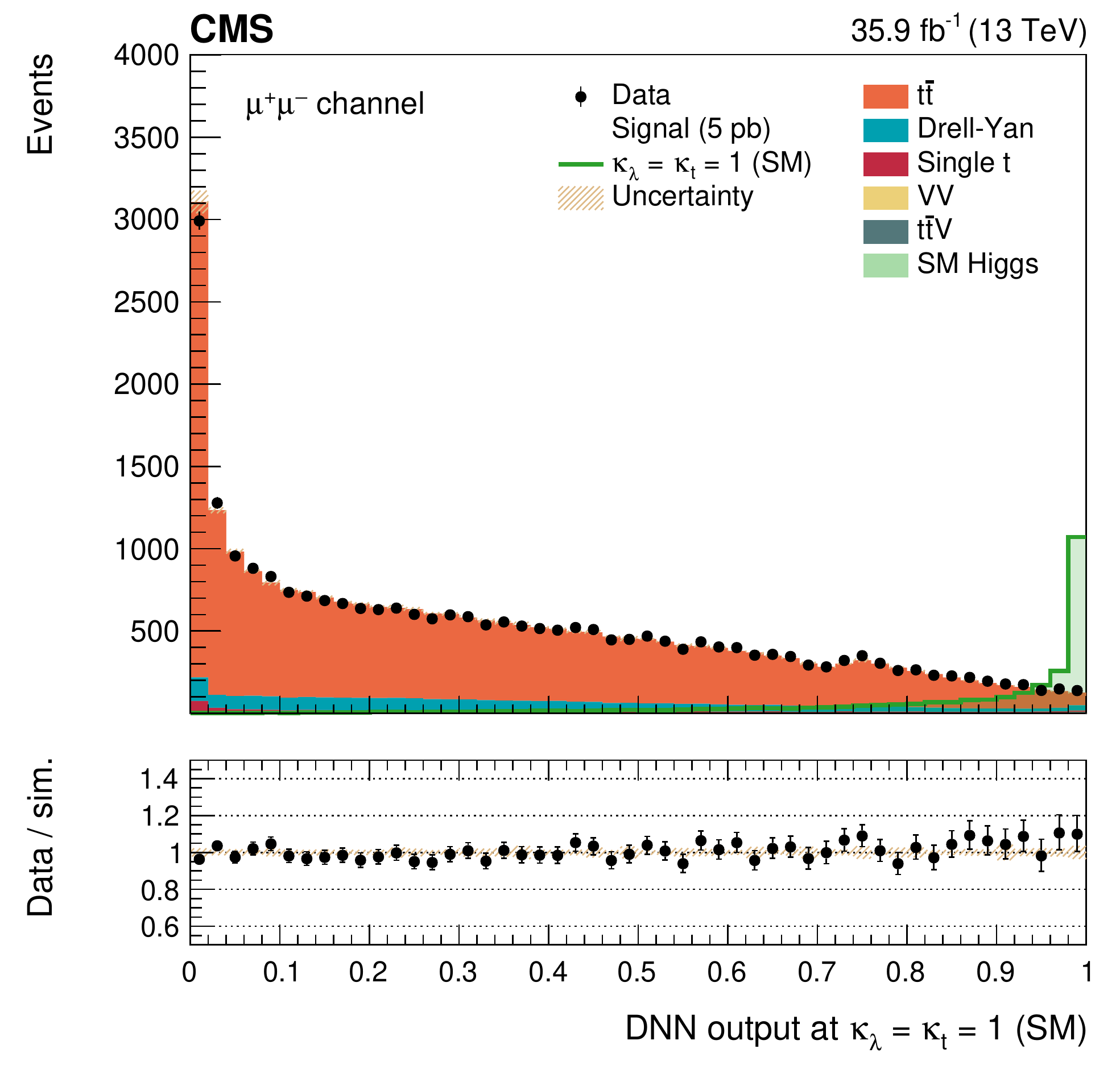}
    \caption{
      The DNN output distributions in data and simulated events after requiring all selection criteria, in the $\Pep\Pem$ (top), $\Pepm{}\PGmmp$ (middle), and $\PGmp\PGmm$ (bottom) channels.
      Output values towards 0 are background-like, while output values towards 1 are signal-like.
      The parameterised resonant DNN output (left) is evaluated at $\mx = 400$\GeV and the parameterised nonresonant
      DNN output (right) is evaluated at $\kappa_{\lambda} =1$, $\kappa_{\PQt} =1$.
      The two signal hypotheses displayed have been scaled to a
      cross section of 5\unit{pb} for display purposes. Error bars indicate statistical uncertainties, while shaded bands show post-fit systematic uncertainties.
      }
    \label{fig:output_DNN}
\end{figure}

\begin{figure}[!htb]
  \centering
    \includegraphics[width=0.45\textwidth]{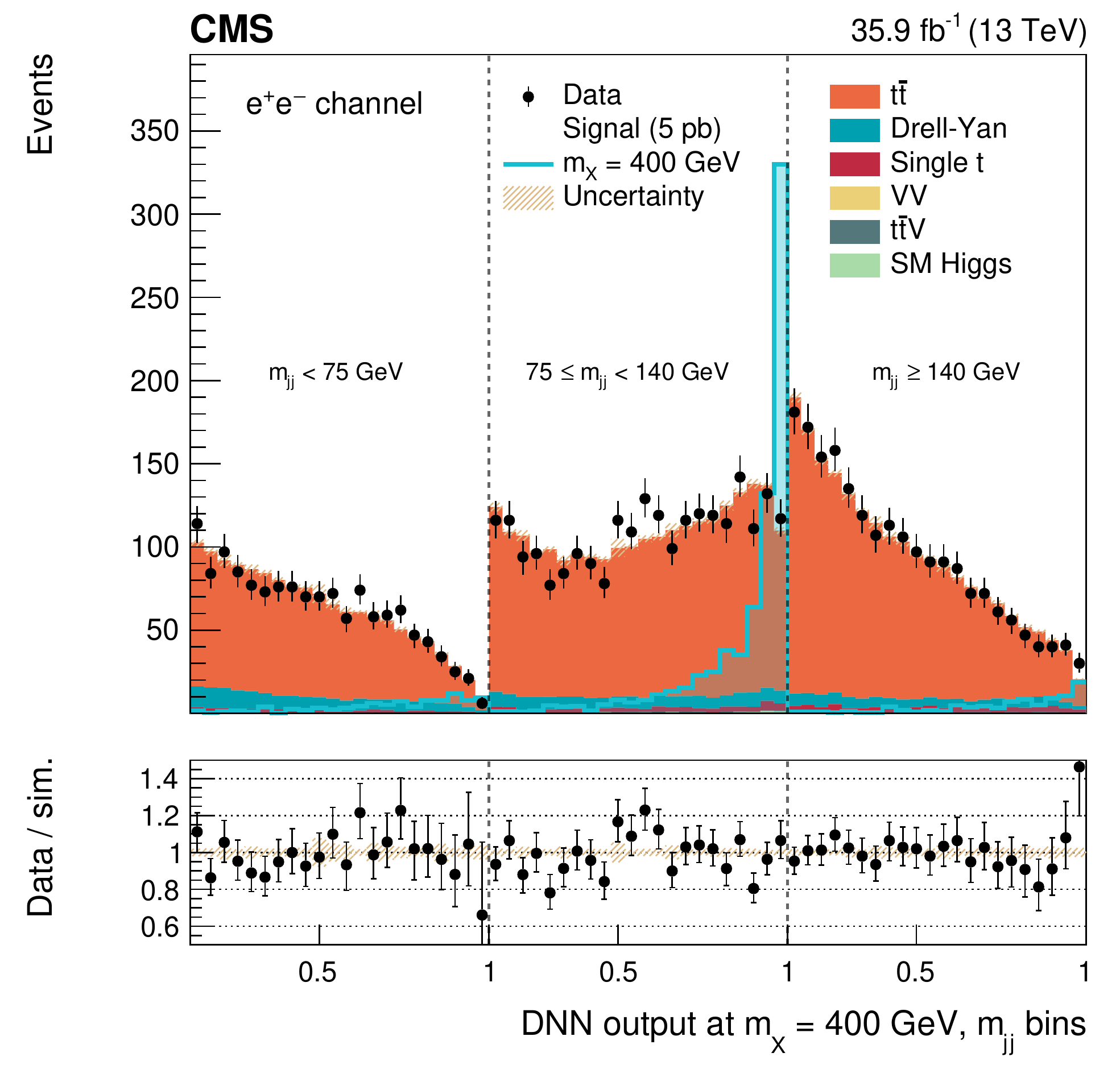}
    \includegraphics[width=0.45\textwidth]{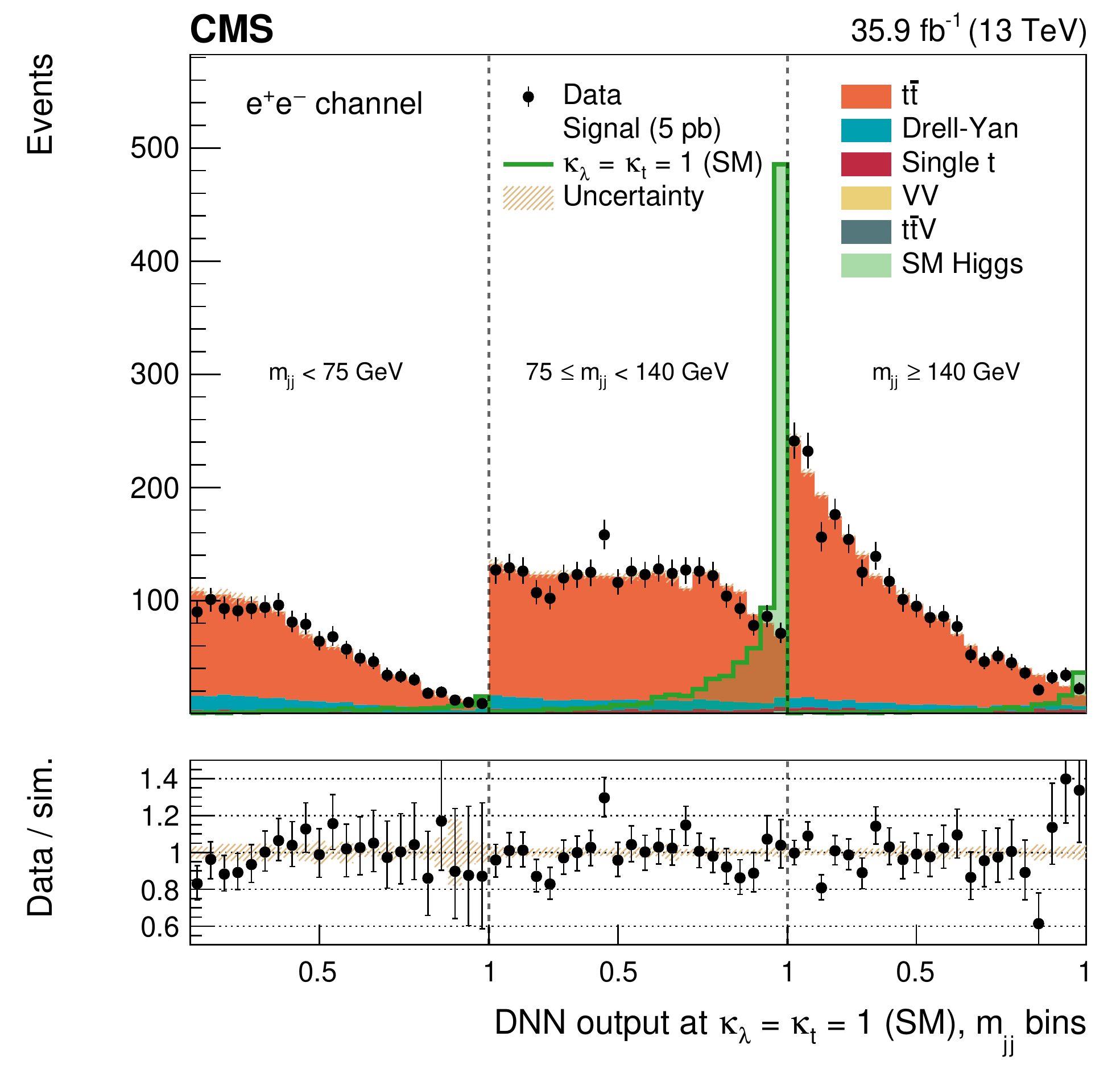}
    \includegraphics[width=0.45\textwidth]{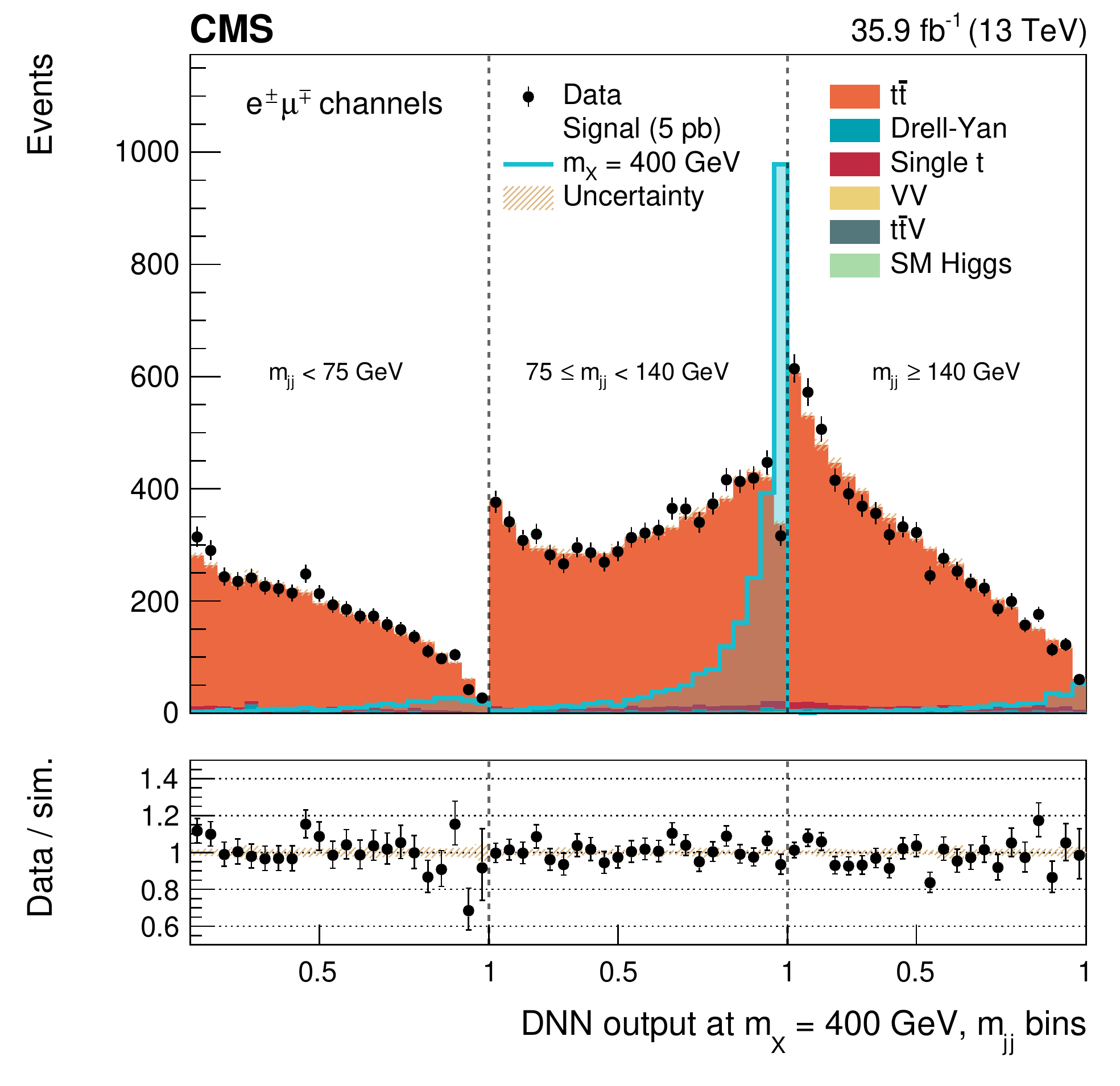}
    \includegraphics[width=0.45\textwidth]{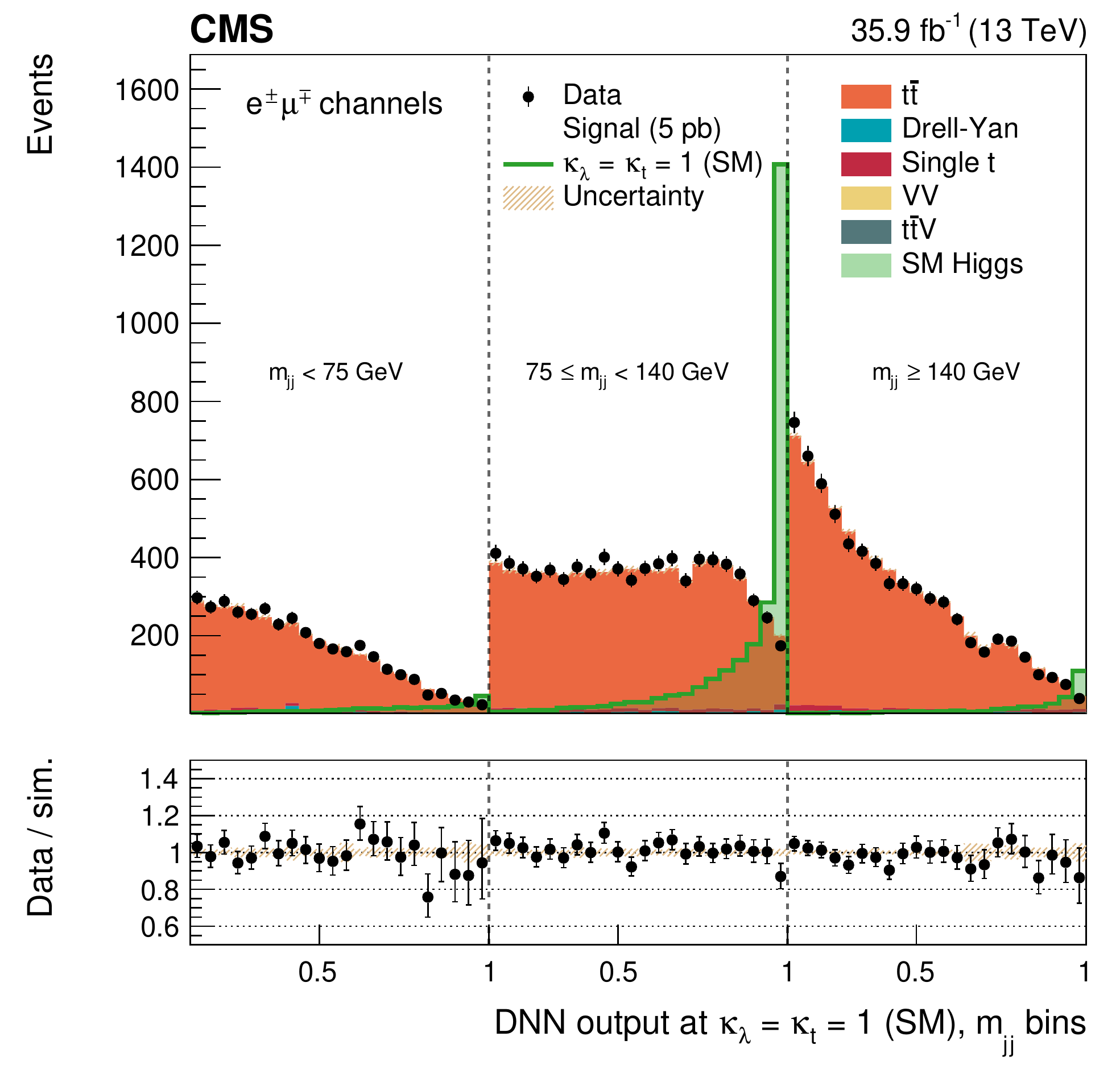}
    \includegraphics[width=0.45\textwidth]{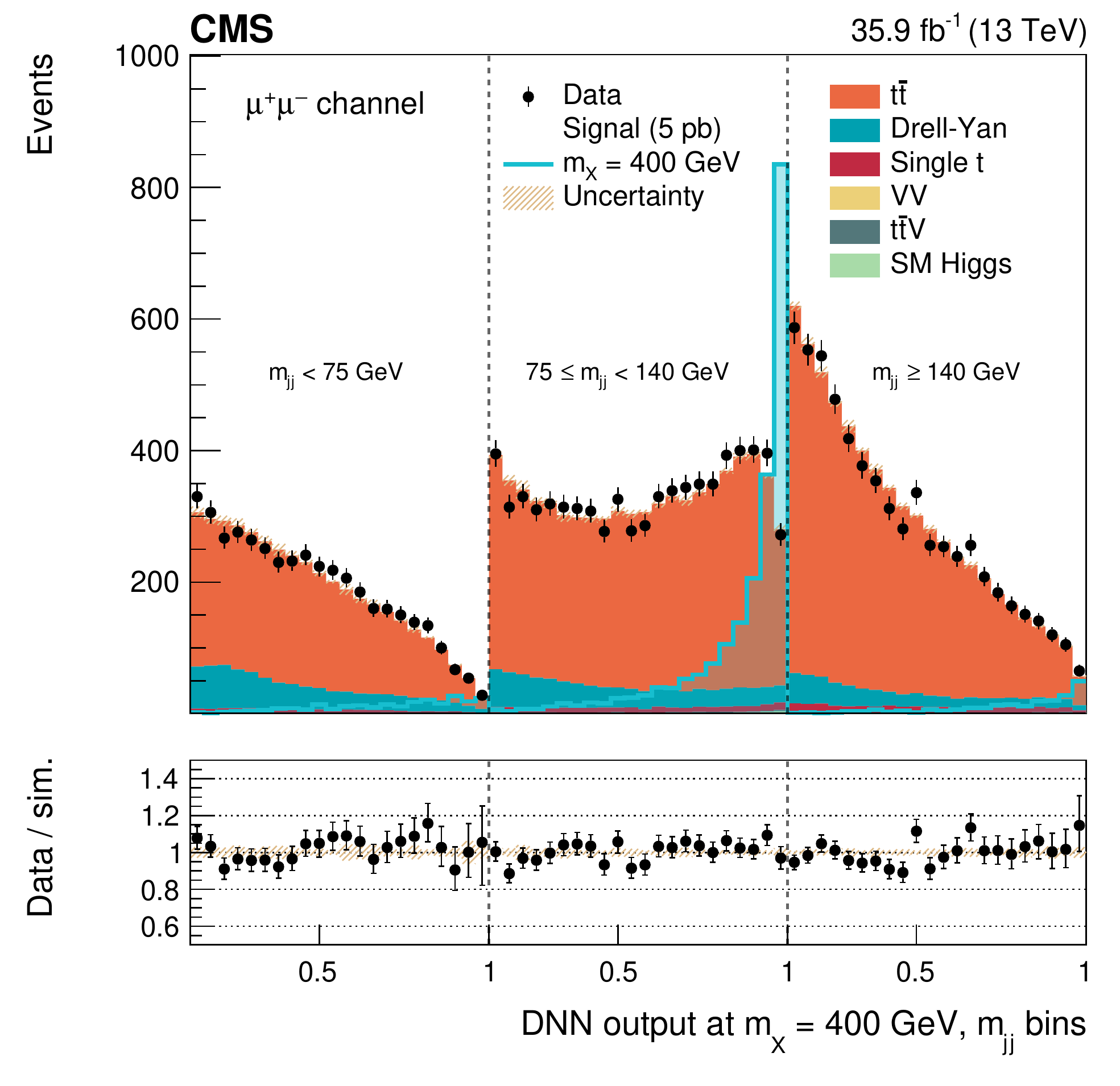}
    \includegraphics[width=0.45\textwidth]{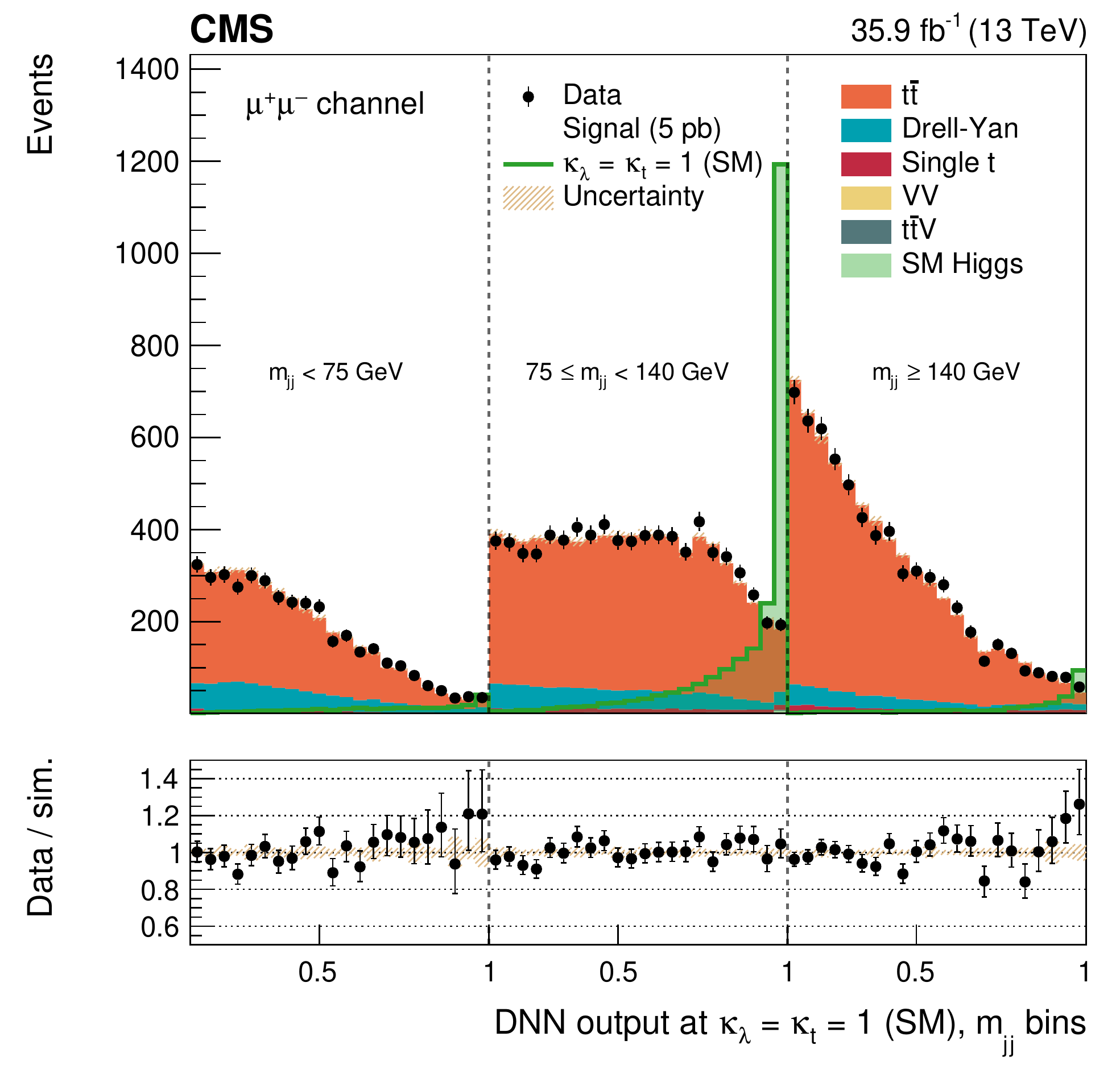}
    \caption{
    The DNN output distributions in data and simulated events, for the $\Pep\Pem$ (top), $\Pepm{}\PGmmp$ (middle), and $\PGmp\PGmm$ (bottom) channels, in three different \mjj
      regions: $\mjj < 75$\GeV, $\mjj \in \left[ \, 75, 140 \, \right)$\GeV, and $\mjj \geq 140$\GeV.
    The parameterised resonant DNN output (left) is evaluated at $\mx = 400$\GeV and the parameterised nonresonant
    DNN output (right) is evaluated at $\kappa_{\lambda} =1$, $\kappa_{\PQt} = 1$.
    The two signal hypotheses displayed have been scaled to a
    cross section of 5\unit{pb} for display purposes. Error bars indicate statistical uncertainties, while shaded bands show post-fit systematic uncertainties.
    }
    \label{fig:flat_mjj_vs_DNN}
\end{figure}

The \mjj distributions, and resonant and nonresonant DNN discriminators
after selection requirements, are shown in Figs.~\ref{fig:mjj} and~\ref{fig:output_DNN}, respectively.
Given their discrimination power
between signal and background, both variables are used to enhance the sensitivity of the analysis.
We define three regions in \mjj: two of them enriched in background, $\mjj < 75$\GeV and $\mjj \geq 140$\GeV,
and the other enriched in signal, $\mjj \in \left[ \, 75, 140 \, \right)$\GeV. In each region, we use the DNN output
as our final discriminant, as shown in Fig.~\ref{fig:flat_mjj_vs_DNN}, where the three \mjj regions
are represented in a single distribution.

\section{Systematic uncertainties} \label{sec:systematics}

We investigate sources of systematic uncertainties and their impact on the statistical interpretation
of the results by considering both uncertainties in the normalisation of the various
processes in the analysis, as well as those affecting the shapes of the distributions.

Theoretical uncertainties in the cross sections of backgrounds
estimated using simulation are considered as systematic uncertainties
in the yield predictions. The uncertainty in the total integrated
luminosity is determined to be 2.5\%~\cite{CMS-PAS-LUM-17-001}.

The following sources of systematic uncertainties that affect the normalisation and shape of the
templates used in the statistical evaluation are considered:
\begin{itemize}

\item \textbf{Trigger efficiency, lepton identification and isolation:} uncertainties in the measurement, using a ``tag-and-probe'' technique, of trigger efficiencies as well as electron and muon isolation and identification efficiencies, are considered as sources of systematic uncertainties. These are evaluated as a function of lepton $\pt$ and $\eta$, and their effect on the analysis is estimated by
varying the corrections to the efficiencies by $\pm$1~standard deviation.

\item \textbf{Jet energy scale and resolution:} uncertainties in the jet energy
scale are of the order of a few percent and are computed as a function of jet $\pt$ and $\eta$~\cite{Khachatryan:2016kdb}.
A difference in the jet energy
resolution of about 10\% between data and simulation is accounted for by worsening
the jet energy resolution in simulation by $\eta$-dependent factors. The
uncertainty due to these corrections is estimated by a variation of the factors applied by
$\pm$1~standard deviation. Variations of jet energies are propagated to $\ptvecmiss$.

\item \textbf{b tagging:} b tagging efficiency and light-flavour mistag rate corrections and associated uncertainties are determined
as a function of the jet $\pt$~\cite{CMSbtag_Run2}. Their effect on the analysis is estimated by
varying these corrections by $\pm$1~standard deviation.

\item \textbf{Pileup:} the measured total inelastic cross section is varied
by $\pm$5\%~\cite{Aaboud:2016mmw} to produce different expected pileup distributions.

\item \textbf{Renormalisation and factorisation scale uncertainty:} this uncertainty is estimated by varying
the renormalisation ($\mu_{\mathrm{R}}$) and the factorisation ($\mu_{\mathrm{F}}$) scales used during
the generation of the simulated samples independently by factors of 0.5, 1, or 2. Unphysical
cases, where the two scales are at opposite extremes, are not considered.
An envelope is built from the 6 possible combinations by keeping maximum and minimum variations for each bin
of the distributions, and is used as an estimate
of the scale uncertainties for all the background and signal samples.

\item \textbf{PDF uncertainty:} the magnitudes of
the uncertainties related to the PDFs and the variation
of the strong coupling constant for each simulated background and signal
process are obtained using variations of the NNPDF~3.0 set~\cite{Ball:2014uwa}, following the PDF4LHC prescriptions~\cite{Botje:2011sn,Alekhin:2011sk}.

\item \textbf{Simulated sample size:} the finite nature of simulated
  samples is considered as an additional source of systematic
  uncertainty. For each bin of the distributions, one additional
  uncertainty is added, where only the considered bin is altered by $\pm$1~standard deviation, keeping the others at their nominal value.

\item \textbf{DY background estimate from data:} the systematic uncertainties
listed above, which affect the simulation samples, are propagated
to $\epsilon_{k}$ and $F_{kl}$, both computed from simulation. These
uncertainties are then propagated to the weights $W_{\text{sim}}$ and to
the normalisation and shape of the estimated DY background contribution.
The uncertainty due to the finite size of the simulation samples used for the determination of
$\epsilon_{k}$ and $F_{kl}$ is also taken into account.
Since previous measurements~\cite{Chatrchyan:2014dha,Khachatryan:2016iob} have shown
that the flavour composition of DY events with associated jets in data
is compatible with the simulation within scale uncertainties, no extra source
of theoretical uncertainty has been considered for $F_{kl}$. To
account for residual differences between the $\Pep\Pem$ and $\PGmp\PGmm$
channels not taken into account by $F_{kl}$, due to the different requirements on lepton $\pt$, a 5\% uncertainty in the
normalisation of the DY background estimate is added in both channels.

\end{itemize}

The effects of these uncertainties on the total yields in the analysis selection are summarised
in Table~\ref{table:sys}.

\begin{table}[htb]
\topcaption{
  Summary of the systematic uncertainties and their impact
  on total background yields and on the SM and $\mx =
  400$\GeV signal hypotheses in the signal
  region.
}
\label{table:sys}
\centering
\begin{tabular}{lcc} \hline
Source & Background yield variation & Signal yield variation \\
\hline
Electron identification and isolation & 2.0--3.2\% & 1.9--2.9\% \\
Jet b tagging (heavy-flavour jets) & 2.5\% & 2.5--2.7\% \\
Integrated luminosity & 2.5\% & 2.5\% \\
Trigger efficiency & 0.5--1.4\% & 0.4--1.4\% \\
Pileup & 0.3--1.4\% & 0.3--1.5\% \\
Muon identification & 0.4--0.8\% & 0.4--0.7\% \\
PDFs & 0.6--0.7\% & 1.0--1.4\% \\
Jet b tagging (light-flavour jets) & 0.3\% & 0.3--0.4\% \\
Muon isolation & 0.2--0.3\% & 0.1--0.2\% \\
Jet energy scale & $<$0.1--0.3\% & 0.7--1.0\% \\
Jet energy resolution & 0.1\% & $<$0.1\% \\[\cmsTabSkip]
    \multicolumn{3}{c}{Affecting only $\ttbar$ (85.1--95.7\% of the total bkg.)} \\
$\mu_{\mathrm{R}}$ and $\mu_{\mathrm{F}}$ scales   & \multicolumn{2}{c}{12.8--12.9\%} \\
$\ttbar$ cross section & \multicolumn{2}{c}{5.2\%} \\
Simulated sample size   & \multicolumn{2}{c}{$<$0.1\%}\\[\cmsTabSkip]
\multicolumn{3}{c}{Affecting only DY in $\Pepm{}\PGmmp$ channel (0.9\% of the total bkg.)} \\
$\mu_{\mathrm{R}}$ and $\mu_{\mathrm{F}}$ scales & \multicolumn{2}{c}{24.6--24.7\%} \\
Simulated sample size & \multicolumn{2}{c}{7.7--11.6\%} \\
DY cross section & \multicolumn{2}{c}{4.9\%}\\[\cmsTabSkip]
\multicolumn{3}{c}{Affecting only DY estimate from data in same-flavour events (7.1--10.7\% of the total bkg.)} \\
Simulated sample size & \multicolumn{2}{c}{18.8--19.0\%} \\
Normalisation & \multicolumn{2}{c}{5.0\%} \\[\cmsTabSkip]
\multicolumn{3}{c}{Affecting only single top quark (2.5--2.9\% of the total bkg.)} \\
Single $\PQt$ cross section & \multicolumn{2}{c}{7.0\%} \\
Simulated sample size & \multicolumn{2}{c}{$<$0.1--1.0\%} \\
$\mu_{\mathrm{R}}$ and $\mu_{\mathrm{F}}$ scales & \multicolumn{2}{c}{$<$0.1--0.2\%}\\[\cmsTabSkip]
Affecting only signal & SM signal & $m_\text{X} = 400\GeV$ \\
$\mu_{\mathrm{R}}$ and $\mu_{\mathrm{F}}$ scales & 24.2\% & 4.6--4.7\% \\
Simulated sample size & $<$0.1\% & $<$0.1\% \\
\hline
\end{tabular}
\end{table}

\section{Results} \label{sec:results}
A binned maximum likelihood fit is performed in order to extract best fit signal cross sections.
The fit is performed using templates built from the DNN output distributions in the three \mjj regions,
as shown in Fig.~\ref{fig:flat_mjj_vs_DNN}, and in the three channels ($\Pep\Pem$, $\PGmp\PGmm$, and $\Pepm\PGmmp$). The likelihood function is the product of the Poisson likelihoods over all bins of the templates and is given by
\begin{equation*}
 L(\beta_{\text{signal}}, \beta_{k}|\text{data}) = \prod_{i=1}^{N_{\text{bins}}}\frac{\mu_{i}^{n_{i}}\, \re^{-\mu_{i}}}{n_{i}!},
\end{equation*}
where $n_i$ is the number of observed events in bin $i$ and the Poisson mean for bin $i$ is given by
\begin{equation*}
\mu_{i} = \beta_{\text{signal}} \, S_{i} + \sum_{k}\beta_{k}\, T_{k,i},
\end{equation*}
where $k$ denotes all of the considered background processes, $T_{k, i}$ is the bin content of bin $i$ of
the template for process $k$, and $S_i$ is the bin content of bin $i$ of the signal template.
The parameter $\beta_{k}$ is the nuisance parameter for the normalisation of the process $k$, constrained by theoretical uncertainties with a log-normal prior,
and $\beta_{\text{signal}}$ is the signal strength, unconstrained.
For each systematic uncertainty affecting the shape (normalisation) of the templates, a nuisance parameter is introduced with a Gaussian (log-normal) prior.

The best-fit values for all the nuisance parameters, as well as the corresponding post-fit uncertainties, are extracted by
performing a binned maximum likelihood fit, in the background-only hypothesis,
of the \mjj~vs.~DNN output distributions (such as Fig.~\ref{fig:flat_mjj_vs_DNN} left) to the data.
Only nuisance parameters affecting the backgrounds are considered.

\subsection{Resonant production} \label{subsec:results_resonant}

\begin{figure}[!htb]
\centering
\includegraphics[width=0.47\textwidth]{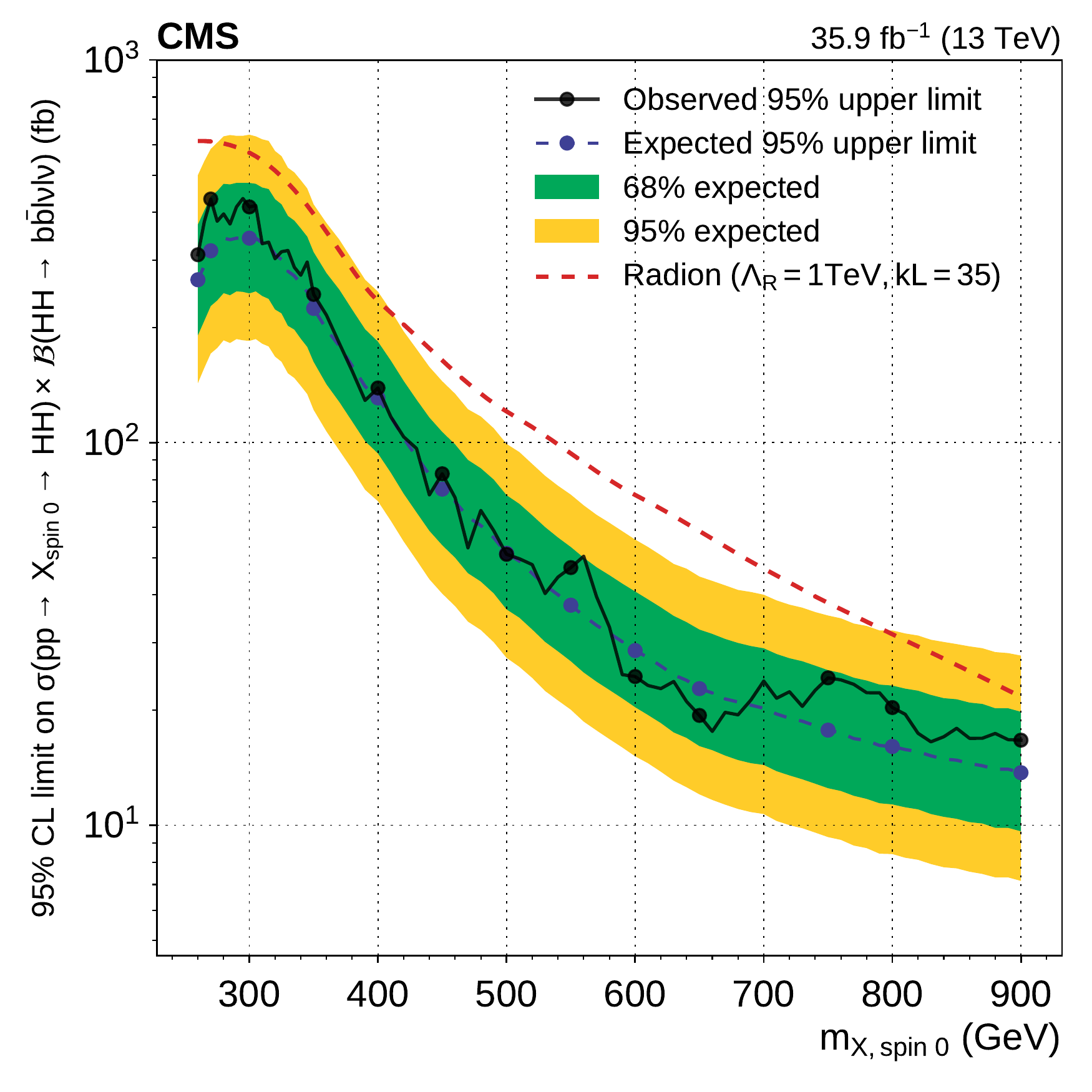}
\includegraphics[width=0.47\textwidth]{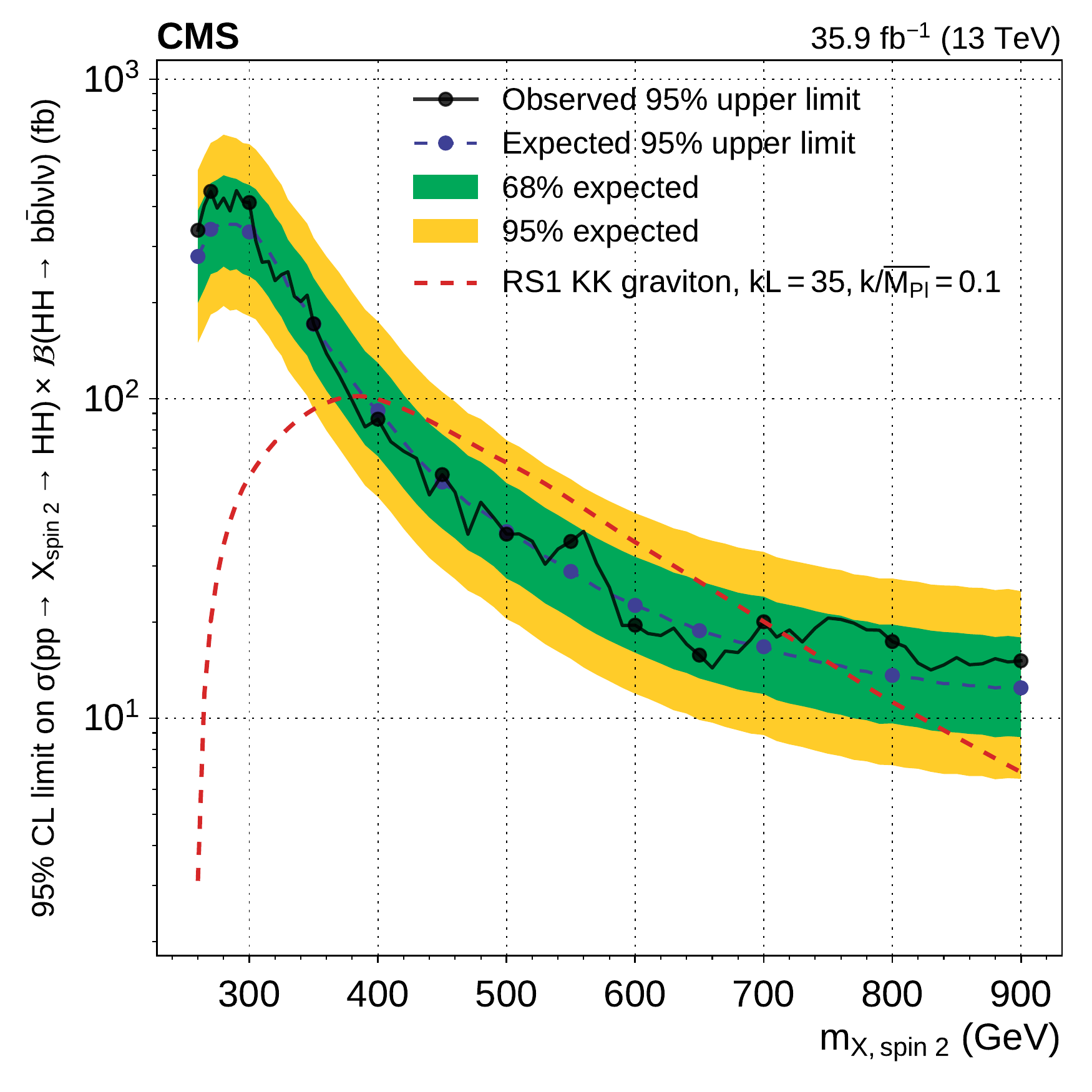}
\caption{
Expected (dashed) and observed (continuous) 95\% CL upper limits on the product of the production cross
section for $\PX$ and branching fraction for $\PX \to \PH\PH \to \bbbar\PV\PV \to
\bbbar\ell\PGn\ell\PGn$, as a function of $\mx$. The inner (green) band and the outer (yellow) band indicate the regions containing 68 and 95\%, respectively, of the distribution of limits expected under the background-only hypothesis. These limits are
computed using the asymptotic $\mathrm{CL_s}$ method, combining the $\Pep\Pem$,
$\PGmp\PGmm$ and $\Pe^{\pm}\PGm^{\mp}$ channels, for spin-0 (left)
and spin-2 (right) hypotheses.
The solid circles represent fully-simulated mass points.
The dashed red lines represent possible cross sections for the production of a radion (left)
or a Kaluza--Klein graviton (right), assuming absence of mixing with the Higgs boson~\cite{Oliveira:2014kla}. Parameters used to compute these cross sections can be found in the legend.
}
\label{fig:reslimits}
\end{figure}

The fit results in signal cross sections compatible with zero;
no significant excess above background predictions is observed for
$\PX$ particle mass hypotheses between 260 and 900\GeV.
We set upper limits at 95\% confidence level (CL)
on the product of the production cross section for $\PX$ and
branching fraction for $\PX \to \PH\PH \to \bbbar\PV\PV \to \bbbar\ell\PGn\ell\PGn$ using the asymptotic
modified frequentist method (asymptotic $\mathrm{CL_s}$)~\cite{Junk:1999kv,LEP-CLs,Cowan:2010js}
as a function of the $\PX$ mass hypothesis. The limits are shown in Fig.~\ref{fig:reslimits}.
The observed upper limits on the product of the production cross
section and branching fraction for a narrow-width spin-0 resonance range from 430 to 17\unit{fb},
in agreement with expected upper limits of $340^{+140}_{-100}$ to $14^{+6}_{-4}$\unit{fb}.
For narrow-width spin-2 particles produced in gluon fusion with minimal gravity-like coupling, the observed upper limits range from 450 to 14\unit{fb},
in agreement with expected upper limits of $360^{+140}_{-100}$ to $13^{+6}_{-4}$\unit{fb}.

The left plot of Fig.~\ref{fig:reslimits} shows possible cross sections for the production of a radion, for the parameters $\Lambda_\mathrm{R} = 1\TeV$ (mass scale) and $\text{kL} = 35$ (size of the extra dimension). The right plot of Fig.~\ref{fig:reslimits} shows possible cross sections for the production of a Kaluza--Klein graviton, for the parameters $\text{k} / \overline{\text{M}_\text{Pl}} = 0.1$ (curvature) and $\text{kL} = 35$. These cross sections are taken from \cite{Oliveira:2014kla}, assuming absence of mixing with the Higgs boson.

\subsection{Nonresonant production}

Likewise for the nonresonant case, the fit results in signal cross sections compatible with zero;
no significant excess above background predictions is seen.
We set upper limits at 95\% CL
on the product of the Higgs boson pair production cross section and branching fraction for
$\PH\PH \to \bbbar\PV\PV \to \bbbar\ell\PGn\ell\PGn$ using the asymptotic $\mathrm{CL_s}$, combining the $\Pep\Pem$, $\PGmp\PGmm$ and $\Pepm\PGmmp$ channels.
The observed upper limit on the SM $\PH\PH \to \bbbar\PV\PV \to \bbbar\ell\PGn\ell\PGn$
cross section is found to be 72\unit{fb}, in agreement with an expected upper limit of $81^{+42}_{-25}$\unit{fb}.
Including theoretical uncertainties in the SM signal cross section,
this observed upper limit amounts to 79 times the SM
prediction, in agreement with an expected upper limit of
$89^{+47}_{-28}$ times the SM prediction.

\begin{figure}[!htb]
  \centering
    \includegraphics[width=0.47\textwidth]{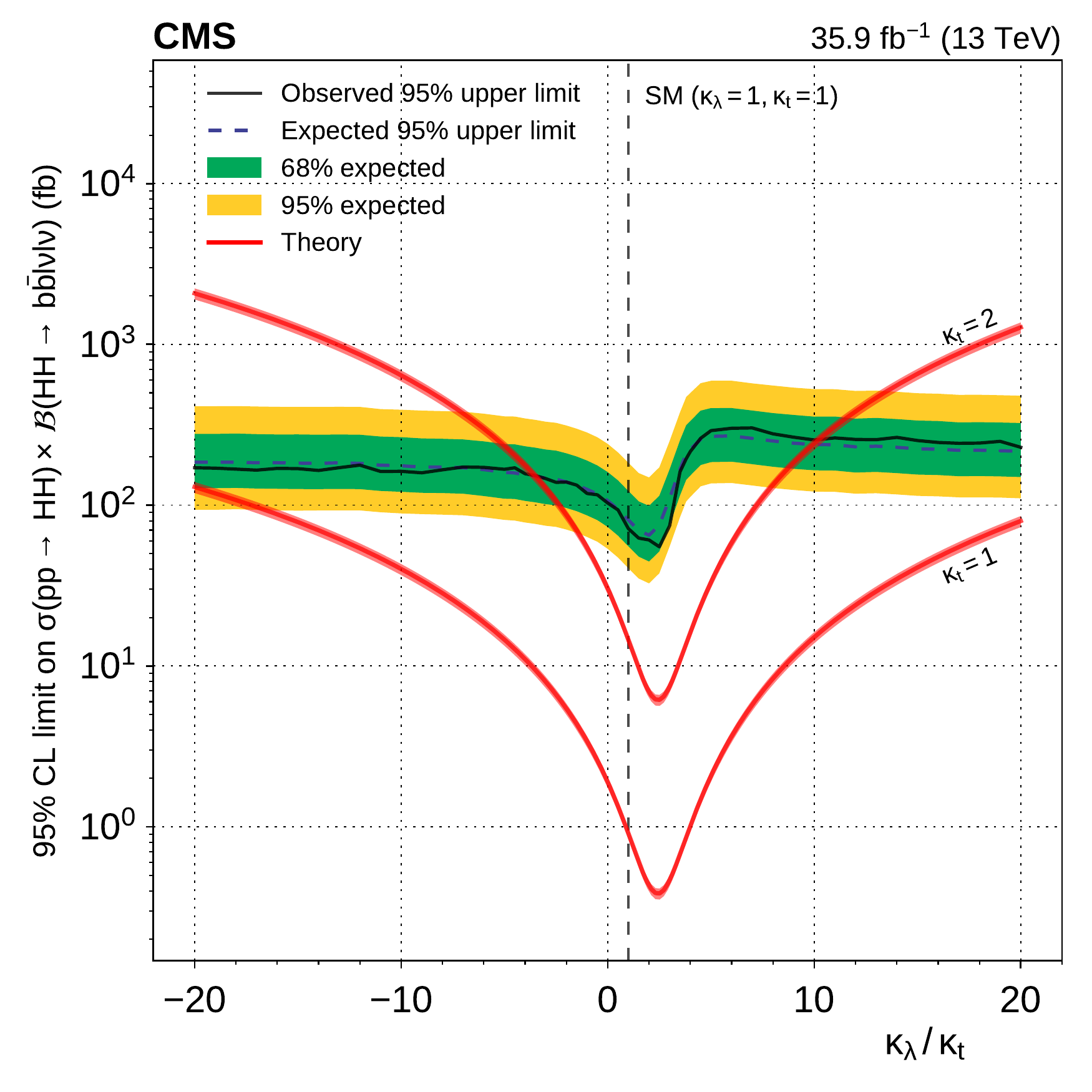}
    \includegraphics[width=0.47\textwidth]{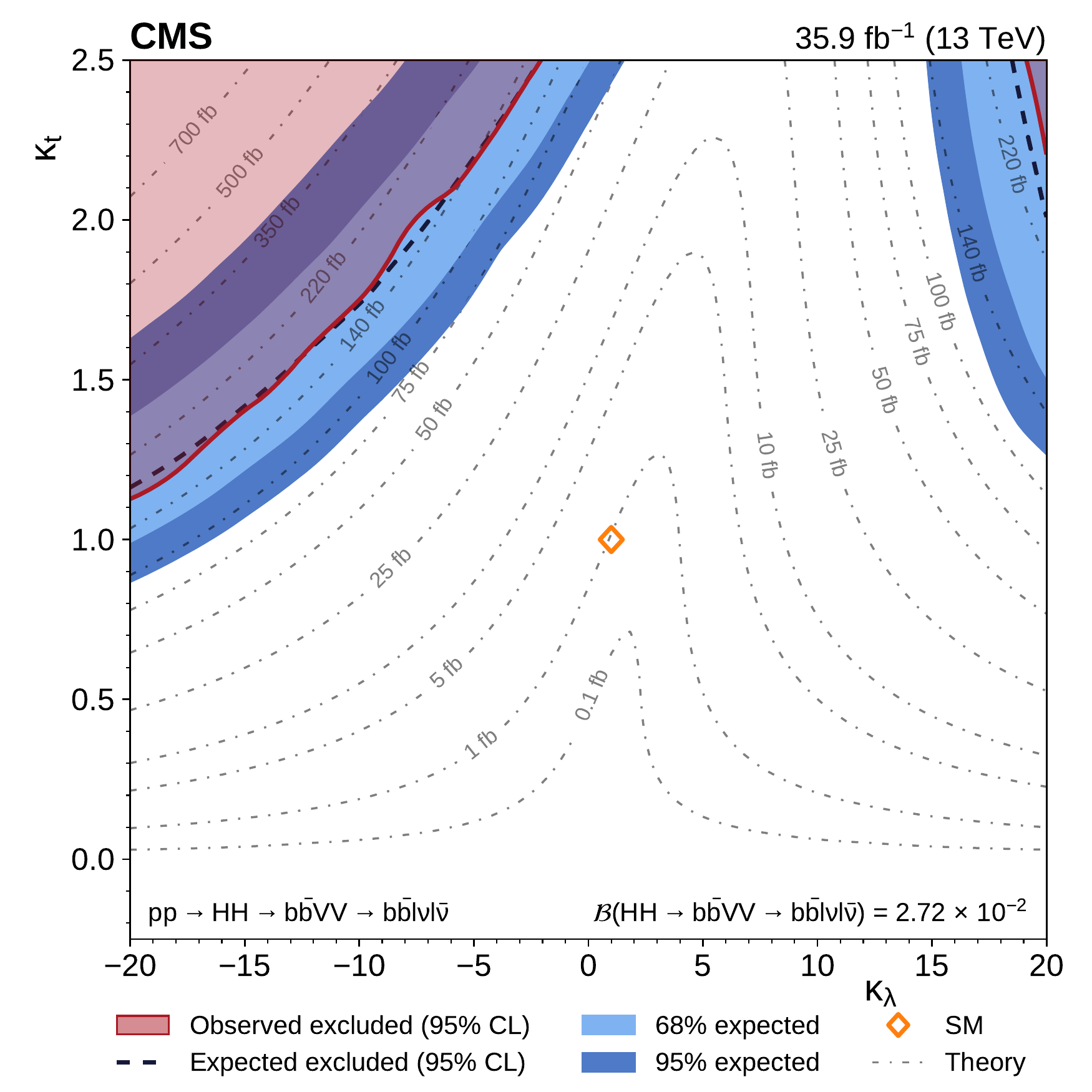}
    \caption{
      Left: expected (dashed) and observed (continuous) 95\% CL upper limits on the product of the Higgs boson pair production cross section and branching fraction for
      $\PH\PH \to \bbbar\PV\PV \to \bbbar\ell\PGn\ell\PGn$ as a
      function of $\kappa_{\lambda} / \kappa_{\PQt}$. The inner (green) band and the outer (yellow) band indicate the regions containing 68 and 95\%, respectively, of the distribution of limits expected under the background-only hypothesis. Red lines show the theoretical cross sections, along with their uncertainties, for $\kappa_{\PQt} = 1$ (SM) and $\kappa_{\PQt} = 2$.
      Right: exclusions in the ($\kappa_{\lambda}$, $\kappa_{\PQt}$) plane. The red region corresponds to parameters excluded at 95\% CL with the observed data, whereas the dashed black line and the blue areas correspond to the expected exclusions and the 68 and 95\% bands (light and dark respectively).
      Isolines of the product of the theoretical cross section and branching fraction for $\PH\PH \to \bbbar\PV\PV
      \to \bbbar\ell\PGn\ell\PGn$ are shown as dashed-dotted lines. The diamond marker indicates the prediction of the SM.
      All theoretical predictions are extracted
      from Refs.~\cite{deFlorian:2016spz,deFlorian:2015moa,deFlorian:2013jea,Borowka:2016ehy,Dawson:1998py,Grigo:2014jma,cluster_hh_lhchxswg}.
    }
    \label{fig:nonreslimits}
\end{figure}

In the BSM hypothesis, upper limits are set as a function
of $\kappa_{\lambda} / \kappa_{\PQt}$, as shown in
Fig.~\ref{fig:nonreslimits} (left panel), since the signal kinematics depend
only on this ratio of couplings.
Red lines show the theoretical cross sections, along with their uncertainties, for $\kappa_{\PQt} = 1$ (SM) and $\kappa_{\PQt} = 2$. The theoretical signal cross section is minimal for $\kappa_{\lambda} / \kappa_{\PQt}=2.45$~\cite{cluster_hh_lhchxswg},
corresponding to a maximal interference between the diagrams shown on Fig.~\ref{fig:diagrams}.

Excluded regions in the $\kappa_{\PQt}$ vs. $\kappa_{\lambda}$ plane are shown
in Fig.~\ref{fig:nonreslimits} (right panel).
The signal cross sections and kinematics are invariant under a
$(\kappa_{\lambda} , \kappa_{\PQt}) \leftrightarrow (-\kappa_{\lambda} , -\kappa_{\PQt})$ transformation,
hence the expected and observed limits on the production cross section,
as well as the constraints on the $\kappa_{\lambda}$ and $\kappa_{\PQt}$ parameters respect the same symmetry.
The red region in the panel corresponds to parameters excluded at 95\% CL with the observed data, whereas the dashed black line and the blue areas correspond to the expected exclusions and the 68 and 95\% bands. Isolines of the product of the theoretical cross section and branching fraction for $\PH\PH \to \bbbar\PV\PV \to \bbbar\ell\PGn\ell\PGn$ are shown as dashed-dotted lines.

\clearpage
\section{Summary}

A search for resonant and nonresonant Higgs boson pair production ($\PH\PH$) is presented,
where one of the Higgs bosons decays to $\bbbar$, and the
other to $\PV\PV \to \ell\PGn \ell\PGn$,  where $\PV$ is either a $\PW$ or a $\PZ$ boson. The LHC proton-proton collision
data at $\sqrt{s}=13$\TeV collected by the CMS experiment corresponding to an integrated
luminosity of 35.9\fbinv are used. Masses are considered in the range
between 260 and 900\GeV for the resonant search, while anomalous Higgs boson
self-coupling and coupling to the top quark are considered in
addition to the standard model case for the nonresonant search.

The results obtained are in agreement, within uncertainties, with the
predictions of the standard model. For the resonant search, the data
exclude a product of the production cross section and branching
fraction of narrow-width spin-0 particles from 430 to 17\unit{fb}, in
agreement with the expectations of $340^{+140}_{-100}$ to
$14^{+6}_{-4}$\unit{fb}, and narrow-width spin-2 particles produced with
minimal gravity-like coupling from 450 to 14\unit{fb}, in agreement with
the expectations of $360^{+140}_{-100}$ to $13^{+6}_{-4}$\unit{fb}. For the
standard model $\PH\PH$ hypothesis, the data exclude a product of the
production cross section and branching fraction of 72\unit{fb},
corresponding to 79 times the SM cross section. The expected
exclusion is $81^{+42}_{-25}$\unit{fb}, corresponding to $89^{+47}_{-28}$
times the SM cross section.

\clearpage

\begin{acknowledgments}
We congratulate our colleagues in the CERN accelerator departments for the excellent performance of the LHC and thank the technical and administrative staffs at CERN and at other CMS institutes for their contributions to the success of the CMS effort. In addition, we gratefully acknowledge the computing centres and personnel of the Worldwide LHC Computing Grid for delivering so effectively the computing infrastructure essential to our analyses. Finally, we acknowledge the enduring support for the construction and operation of the LHC and the CMS detector provided by the following funding agencies: BMWFW and FWF (Austria); FNRS and FWO (Belgium); CNPq, CAPES, FAPERJ, and FAPESP (Brazil); MES (Bulgaria); CERN; CAS, MoST, and NSFC (China); COLCIENCIAS (Colombia); MSES and CSF (Croatia); RPF (Cyprus); SENESCYT (Ecuador); MoER, ERC IUT, and ERDF (Estonia); Academy of Finland, MEC, and HIP (Finland); CEA and CNRS/IN2P3 (France); BMBF, DFG, and HGF (Germany); GSRT (Greece); OTKA and NIH (Hungary); DAE and DST (India); IPM (Iran); SFI (Ireland); INFN (Italy); MSIP and NRF (Republic of Korea); LAS (Lithuania); MOE and UM (Malaysia); BUAP, CINVESTAV, CONACYT, LNS, SEP, and UASLP-FAI (Mexico); MBIE (New Zealand); PAEC (Pakistan); MSHE and NSC (Poland); FCT (Portugal); JINR (Dubna); MON, RosAtom, RAS, RFBR and RAEP (Russia); MESTD (Serbia); SEIDI, CPAN, PCTI and FEDER (Spain); Swiss Funding Agencies (Switzerland); MST (Taipei); ThEPCenter, IPST, STAR, and NSTDA (Thailand); TUBITAK and TAEK (Turkey); NASU and SFFR (Ukraine); STFC (United Kingdom); DOE and NSF (USA).

\hyphenation{Rachada-pisek} Individuals have received support from the Marie-Curie programme and the European Research Council and Horizon 2020 Grant, contract No. 675440 (European Union); the Leventis Foundation; the A. P. Sloan Foundation; the Alexander von Humboldt Foundation; the Belgian Federal Science Policy Office; the Fonds pour la Formation \`a la Recherche dans l'Industrie et dans l'Agriculture (FRIA-Belgium); the Agentschap voor Innovatie door Wetenschap en Technologie (IWT-Belgium); the Ministry of Education, Youth and Sports (MEYS) of the Czech Republic; the Council of Science and Industrial Research, India; the HOMING PLUS programme of the Foundation for Polish Science, cofinanced from European Union, Regional Development Fund, the Mobility Plus programme of the Ministry of Science and Higher Education, the National Science Center (Poland), contracts Harmonia 2014/14/M/ST2/00428, Opus 2014/13/B/ST2/02543, 2014/15/B/ST2/03998, and 2015/19/B/ST2/02861, Sonata-bis 2012/07/E/ST2/01406; the National Priorities Research Program by Qatar National Research Fund; the Programa Clar\'in-COFUND del Principado de Asturias; the Thalis and Aristeia programmes cofinanced by EU-ESF and the Greek NSRF; the Rachadapisek Sompot Fund for Postdoctoral Fellowship, Chulalongkorn University and the Chulalongkorn Academic into Its 2nd Century Project Advancement Project (Thailand); and the Welch Foundation, contract C-1845. \end{acknowledgments}

\bibliography{auto_generated}
\cleardoublepage \appendix\section{The CMS Collaboration \label{app:collab}}\begin{sloppypar}\hyphenpenalty=5000\widowpenalty=500\clubpenalty=5000\textbf{Yerevan Physics Institute,  Yerevan,  Armenia}\\*[0pt]
A.M.~Sirunyan, A.~Tumasyan
\vskip\cmsinstskip
\textbf{Institut f\"{u}r Hochenergiephysik,  Wien,  Austria}\\*[0pt]
W.~Adam, F.~Ambrogi, E.~Asilar, T.~Bergauer, J.~Brandstetter, E.~Brondolin, M.~Dragicevic, J.~Er\"{o}, M.~Flechl, M.~Friedl, R.~Fr\"{u}hwirth\cmsAuthorMark{1}, V.M.~Ghete, J.~Grossmann, J.~Hrubec, M.~Jeitler\cmsAuthorMark{1}, A.~K\"{o}nig, N.~Krammer, I.~Kr\"{a}tschmer, D.~Liko, T.~Madlener, I.~Mikulec, E.~Pree, D.~Rabady, N.~Rad, H.~Rohringer, J.~Schieck\cmsAuthorMark{1}, R.~Sch\"{o}fbeck, M.~Spanring, D.~Spitzbart, W.~Waltenberger, J.~Wittmann, C.-E.~Wulz\cmsAuthorMark{1}, M.~Zarucki
% ...
\vskip\cmsinstskip
% ...
1:~~Also at Vienna University of Technology, Vienna, Austria\\
% ...

\end{sloppypar}
\end{document}